\documentclass[journal]{IEEEtran}
\usepackage{graphicx}
\usepackage{amssymb}
\usepackage{amsmath}
\usepackage{amssymb,amsfonts}
\usepackage{algorithm}
\usepackage{algorithmic}
\usepackage{booktabs}
\usepackage{ragged2e}
\usepackage{multirow} 
\usepackage{threeparttable}
\usepackage{cite}
\usepackage{textcomp}
\usepackage{xcolor}
\usepackage{stfloats}
\usepackage{booktabs}  
\usepackage{threeparttable}
\usepackage{multirow} 
\usepackage{epsfig}
\usepackage{threeparttable}
\usepackage{verbatim}
\usepackage{subfigure}
\usepackage{graphicx}
\usepackage{float}
\usepackage{color}
\usepackage{flushend}
\usepackage{url}

\ifCLASSINFOpdf
\else
\fi
\begin{document}
%
\title{MDENet: Multi-modal Dual-embedding Networks for Malware Open-set Recognition}
%
%
%

\author{Jingcai~Guo,  
       Yuanyuan~Xu, 
       Wenchao~Xu,
       Yufeng~Zhan,
       Yuxia~Sun, 
	and~Song~Guo
\thanks{J. Guo, W. Xu and S. Guo are with Department of Computing, The Hong Kong Polytechnic University, Hong Kong SAR, China, and with The Hong Kong Polytechnic University Shenzhen Research Institute, Shenzhen 518057, China.}
\thanks{Y. Xu is with the School of Information and Software Engineering, University of Electronic Science and Technology of China, Chengdu 610054, China.}
\thanks{Y. Zhan is with the School of Automation, Beijing Institute of Technology, Beijing 100081, China.}
\thanks{Y. Sun is with the Department of Computer Science, Jinan University, Guangzhou 510632, China, and Guangdong Provincial Key Laboratory of High-Performance Computing, Guangzhou 510275, China.}
}
\maketitle

\begin{abstract}
Malware open-set recognition (MOSR) aims at jointly classifying malware samples from known families and detect the ones from novel unknown families, respectively. Existing works mostly rely on a \textit{well-trained} classifier considering the predicted probabilities of each known family with a threshold-based detection to achieve the MOSR. 
However, our observation reveals that the feature distributions of malware samples are extremely similar to each other even between known and unknown families. Thus the obtained classifier may produce overly high probabilities of testing unknown samples toward known families and degrade the model performance. 
In this paper, we propose the Multi-modal Dual-Embedding Networks, dubbed \textit{MDENet}, to take advantage of comprehensive malware features from different modalities to enhance the diversity of malware feature space, which is more representative and discriminative for down-stream recognition. 
Concretely, we first generate a malware image for each observed sample based on their numeric features using our proposed numeric encoder with a re-designed multi-scale CNN structure, which can better explore their statistical and spatial correlations. 
Besides, we propose to organize tokenized malware features into a sentence for each sample considering its behaviors and dynamics, and utilize language models as the textual encoder to transform it into a representable and computable textual vector. 
Such parallel multi-modal encoders can fuse the above two components to enhance the feature diversity. 
Last, to further guarantee the open-set recognition, we dually embed the fused multi-modal representation into one primary space and an associated sub-space, i.e., discriminative and exclusive spaces, with contrastive sampling and $\rho$-bounded enclosing sphere regularizations, which resort to classification and detection, respectively. 
Moreover, we also enrich our previously proposed large-scaled malware dataset \textit{MAL-100} with multi-modal characteristics and contribute an improved version dubbed \textit{MAL-100$^{+}$}. Experimental results on the widely used malware dataset \textit{Mailing} and the proposed \textit{MAL-100$^{+}$} demonstrate the effectiveness of our method.
\end{abstract}


%
\IEEEpeerreviewmaketitle

\section{Introduction}
\label{sec:introduction}
%
%
%
%

\begin{figure}[t]
	\centering
	\includegraphics[width=0.44\textwidth]{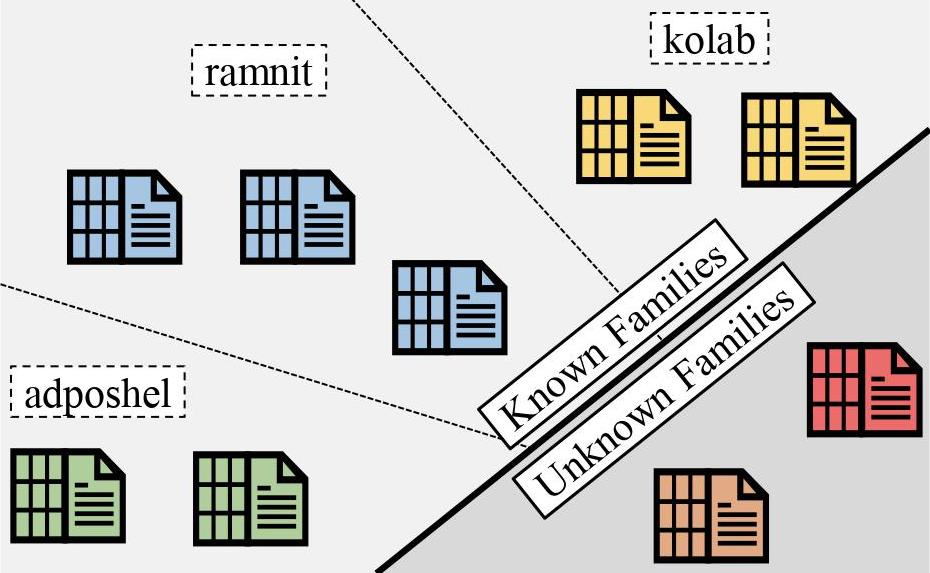}
	\caption{Malware open-set recognition: the samples of only \textit{known} families are presented during training, while the recognition system is expected to recognize samples from both \textit{known} and \textit{unknown} families, respectively (better viewed in color).} 
	\label{fig: osr}
\end{figure}

Machine deep learning has made tremendous success in the past few years across multiple real-world applications~\cite{deng2009imagenet,batista2004study,lin2011machine,guo2020novel,lv2014traffic,nguyen2015topic,guo2020dual,lian2018xdeepfm,he2016deep,simonyan2014very,guo2019adaptive,yang2015facial,shone2018deep,mahapatra2021medical,guo2023graph,liu2022towards,guo2021conservative,lu2023decomposed,wang2023data,liu2023zsl,guo2023application,huo2023offline,zhou2021device,guo2021learning,wang2022exploring,ma2019position,guo2016improved,huo2022procc,zhou2022cadm,guo2019ams,wang2022efficient,guo2019ee,zhou2021octo,guo2022fed}. In recent years, malicious software (a.k.a. malware) such as computer viruses, worms, trojan horses, spywares, etc., can cause severe threats to computer and many more device systems. Recently, as the explosive increasing of malware samples over the years, more and more challenges have emerged and brought about many security and operational issues in cyber-security~\cite{rieck2008learning,zhou2012dissecting} and many more related fields such as data sharing and retrieval~\cite{cavallaro2020deep,janani2021secure}, malicious image and voice attacks~\cite{zhang2020voiceprint,zhang2020robust, zhou2021face}, secure distributed system~\cite{cao2021provably} and so on. 
The malware recognition is an emerging research topic that aims at classifying malware samples into different families, in which the samples of the same family are usually equipped with similar attack techniques. Further investigations and precautionary measures can then be applied to protect various device systems, proactively. 
In recent years, the malware recognition, i.e., \textit{classification particularly}, has obtained fairly promising results with low error rates~\cite{dahl2013large,kong2013discriminant,pascanu2015malware,huang2016mtnet,ni2018malware,vasan2020imcfn}. However, it is noticed that the conventional malware recognition system usually holds a relatively strong assumption that all malware families are known to the system in a \textit{close-set} scenario, i.e., the recognition model is trained and tested within the same category of malware families. Such a close-set setting is acceptable for a long period of time because on the one hand, the malware recognition system is usually implemented in a certain range of application areas, where the families are usually relatively stable and sealed. On the other hand, it is also impossible to fully collect all malware families from the whole cyber-network. 

However, with the increasing popularity of various applications and growth of internet users, more and more malware samples are constantly released by malware attackers. It is sourced from NortonLifeLock\footnote{https://www.nortonlifelock.com/} that, over 317 million novel malware samples are produced annually, and many of them are not belong to any malware family we have known before~\cite{NortonLifeLock}. 
Hence, facing the increasing malware samples, it is probable to observe malware samples from novel unknown families that have never been encountered during training. Since most existing malware classification models neglect unknown families with a close-set setting, these malware samples of novel unknown families are thus usually \textit{forcedly} recognized as one of the known families. As a result, it affects the classification performance and many more following down-stream analysis. 
%
It's worth noting that, although several malware \textit{novelty detection} models, e.g.,~\cite{pang2021anomaly}, focus on identifying if a malware sample belongs to known families or not, i.e., as a binary classification, they neglect the accurate \textit{family-level} classification. Thus, it is of crucial importance to correctly identify/detect given malware samples as either known or unknown families, and simultaneously classify those samples of known families into their corresponding families as accurately as possible (Fig.~\ref{fig: osr}). This problem setting can be considered as the \textit{malware open-set recognition} (MOSR) which has been non-extensively investigated before. 

Apart from the malware recognition, the concept of \textit{open-set} was first investigated in computer vision domain~\cite{scheirer2012toward}. For example, there have been several image open-set recognition approaches that designed the threshold-based method relying on the recognition probabilities of a \textit{well-trained} classifier~\cite{bendale2016towards,ge2017generative,neal2018open,liu2019large}. Specifically, a properly defined threshold is first compared with the predicted probabilities of known classes, e.g., in the form of softmax in neural network, to identify the \textit{known/unknown} classes, and the maximum probability indicates the corresponding known class, if identified as \textit{known} previously. Most recently, a few number of malware recognition models, e.g.,~\cite{hassen2018learning,cordonsky2018deeporigin,jordaney2017transcend,guo2021conservative}, followed such a \textit{probability-based} manner to \textit{partially} implement the open-set recognition on malware domain and obtained some decent results. 
%
%
Despite the progress made, we observe that the existing MOSR still faces some challenges that hinder its further development: 1) First, the divergence of the characteristic features of malware samples is \textit{far less} than that of images, and can thus result in large overlaps among different malware families even between known and unknown ones. This difference may encourage the classifier to produce overly high recognition probability on all malware samples and makes the detection degrades dramatically. 2) Second, existing methods only utilize the extracted numeric features, e.g., statical characteristics as an arranged form such as image, to train an \textit{off-the-shelf} classifier, while neglect making use of comprehensive multi-modal malware features to better fuse the malware feature space. 3) Third, there is no large and \textit{multi-modal-based} open-set malware dataset for MOSR setting.

To address the above challenges and construct a more robust MOSR system, we propose the Multi-modal Dual-Embedding Networks, dubbed \textit{MDENet}, to fully exploit and fuse comprehensive multi-modal malware features to the enhanced feature space that is endowed with more diversities to facilitate the training of MOSR classifier. 
Specifically, we apply \textit{Ember}~\cite{anderson2018ember}, i.e., a widely used tool for malware feature extraction, to extract static malware features in both numeric and textual (tokenized) modalities. 
As to the numeric features, we generate a malware image for each sample and make use of a re-designed multi-scale CNN architecture as the numeric encoder to better explore the statistical and spatial correlations. 
%
Within the numeric encoder, we integrate the classic non-local means~\cite{buades2005non} into a re-designed CNN architecture that enables local and global receptive fields in a multi-scale manner. Thus, the numeric encoder can be endowed with the ability to extract multi-correlations of malware features with nearby and long-distance dependence for better representation.
On the other hand, we further generate a sentence for the same sample based on the its textual features. In such a sentence, words can explicitly represent \textit{import} domains of behaviors and dynamics including libraries, modules, functions, etc., used in this malware sample. A language model, e.g., \textit{BERT}~\cite{devlin2018bert} or its variants, is applied as the textual encoder to obtain the representation vector, which will be further fused with the one of the numeric features. 
To construct more discriminative representation for both classification and detection, we take a further step to make use of contrastive sampling and $\rho$-bounded enclosing sphere regularizations, to modify the output of the multi-modal encoders into one primary space and an associated sub-space, namely, discriminative and exclusive spaces, to jointly guarantee the performance of both tasks.

Our contributions are summarized as follows:
\begin{itemize}
\item We formally and practically investigate and analyze the malware open-set recognition problem on large-scale multi-modal malware dataset for better recognition performance.
\item We propose a novel and robust malware open-set recognition framework involving the multi-modal encoders and dual-embedding space learning approach to fuse multi-modal malware features, i.e., numeric and textual (tokenized) features, for better recognition in both classification and detection. We also design a modified distance-based detection mechanism to address the detection degradation in conventional methods.
\item We propose a new version of large-scale malware dataset, dubbed \textit{MAL-100$^{+}$}, covering both conventional extracted numeric features and textual (tokenized) features, which is complementary to our previously proposed \textit{MAL-100}.
\item Experimental results on the widely used malware dataset \textit{Mailing} and our \textit{MAL-100$^{+}$} demonstrate the effectiveness of our method.
\end{itemize}

\section{Related Work}
\label{sec:related_work}

\subsection{Malware Family Recognition (MFR)}
\label{subsec:malware_family_recognition}

\subsubsection{Family-wise Classification for Malware Recognition}
\label{subsubsec:family-wise_classification_for_malware_recognition}
There have been a great number of studies focusing on conventional close-set malware recognition that aims at correctly classifying malware samples into several known malware families. Such a family-wise classification is typically based on malware characteristics collected by analyzing malware statically~\cite{sun2018deep,kalash2018malware} or dynamically~\cite{hansen2016approach,stiborek2018multiple}. Since the dynamic analysis usually requires time-consuming execution of various malware within the sandbox environment, thus the static analysis (i.e., collecting and analyzing some static features such as installation methods, activation mechanisms, natures of carried malicious payloads, etc.) is more appropriate and widely used for real-time malware recognition, particularly for classification. 
Some representative works include~\cite{dahl2013large,kong2013discriminant,wang2014exploring,pascanu2015malware,huang2016mtnet,fan2017dapasa,yerima2019droidfusion,vasan2020imcfn}. For example, 
%
%
Kong \textit{et al.}~\cite{kong2013discriminant} proposed to discriminate malware distance metrics that evaluate the similarity between the attributed function call graphs of two malware programs. 
Wang \textit{et al.}~\cite{wang2014exploring} evaluated the usefulness of risky permissions for malware recognition using SVM, random forest, and decision trees. 
Fan \textit{et al.}~\cite{fan2017dapasa} proposed to make use of sensitive subgraphs to construct five features then fed into several machine learning algorithms such as decision tree, random forest, KNN, and PART for detecting Android malware piggybacked apps in a binary fashion. 
Yerima \textit{et al.}~\cite{yerima2019droidfusion} proposed to train and fuse multi-level classifiers to form a final strong classifier. 
Vasan \textit{et al.}~\cite{vasan2020imcfn} proposed to apply the assemble and fine-tuned CNN architectures to capture more semantic and rich features from malware samples. 
%


\subsubsection{Detection of Unknown Malware Families}
\label{subsec:Detection of Unknown Malware Families}
To the best of our knowledge, there are very few studies related with malware open-set recognition that aims to categorize each malware sample either into its belonging family if it comes from known families, or into a new category which is exclusive from the known ones. Recently, several works considered the novelty detection in malware analysis to explore novel unseen families~\cite{hassen2018learning,cordonsky2018deeporigin,jordaney2017transcend,guo2021conservative}. 
Specifically, 
Hassen \textit{et al.}~\cite{hassen2018learning} build a machine learning-based classification system by combining random-forest classifiers and kernel-density tree-based outlier detectors. However, the system may suffer from scalability issues since one will have to train as many outlier detectors as the number of known malware families. 
Cordonsky \textit{et al.}~\cite{cordonsky2018deeporigin} utilize a sandbox to capture dynamic malware characteristics and train a deep neural network with known malware families. The pre-softmax layer is removed to create a signature generation model, and the Euclidean distance of each point from the origin is used to measure the novelty of a malware family. However, the sandbox experiment is a very time-consuming process and uses an isolation mechanism for malware samples, which makes real-time recognition impossible. 
Jordaney \textit{et al.}~\cite{jordaney2017transcend} use a statistical comparison of samples seen during deployment with those used to train the model, thereby building metrics for prediction quality. However, the detection process only performs well in binary classification case, while degrades dramatically in multiclass case, i.e., 509 out of total 751 unseen malware samples are wrongly classified as known families, which results in imbalanced performance between classification and detection. 
Worse still, none of~\cite{hassen2018learning,cordonsky2018deeporigin,jordaney2017transcend} has paid enough attention to the family-wise classification and haven't been evaluated on large-scale malware datasets to fit a more applicable scenario. 
%
Most recently, our previous work~\cite{guo2021conservative} has first established a formal framework for malware open-set recognition and partially addressed the imbalanced performance by utilizing a conservative novelty synthesizing network in a large-scale malware dataset. However, the limitation still remains. First, the GANs-based synthesizing network usually requires numerous resource in both computing capability and data, which is not fully feasible in every device. Second, only extracted numeric features are considered while ignores comprehensive malware features in different modalities to learn more representative and discriminative feature space. 
%
%

\subsubsection{Our Method}
In this paper, we also follow the main-stream of static analysis as stated in Sect.~\ref{subsubsec:family-wise_classification_for_malware_recognition}. 
Moreover, existing open-source malware datasets are usually limited in size, volume, and modality. For example, only 9 and 25 families are involved in widely used malware datasets \textit{Kaggle2015}~\cite{kaggle2015} and \textit{Mailing}~\cite{nataraj2011malware}, respectively, and to the best of our knowledge, only single modality of malware features is considered previously. We hereby contribute a new large-scale malware dataset, dubbed \textit{MAL-100$^{+}$}, containing 100 families and more than 50 thousand labeled samples along with multi-modal malware features to fill the gap. 
More importantly, our proposed method, i.e., \textit{MDENet}, is more efficient and can construct a more representative and discriminative feature space, thus jointly improves the classification and detection performance.

\subsection{Image Open-set Recognition}
\label{subsec:open-set_recognition}
The open-set recognition (OSR) was first investigated by Scheirer \textit{et al.}~\cite{scheirer2012toward} in the domain of computer vision, where the task is defined to jointly achieve the known class classification and novel unknown class detection. Existing image OSR methods can be grouped into two categories include traditional methods and deep neural networks (DNNs)-based methods.

\subsubsection{Traditional Methods}
The traditional methods are typically implemented by some classic machine learning models such as SVM, nearest-neighbors, sparse-representation, etc., or statistical models such as extreme value machine.  
Among them, the \textit{1-vs-set machine}~\cite{scheirer2012toward} and \textit{W-SVM}~\cite{scheirer2014probability} are the pioneer works and both of them are derived from SVM. The former augments the decision plane of classic SVM from the marginal distances, and the latter integrates useful properties of statistical extreme value theory into SVM to calibrate the classification scores. 
Afterwards, another SVM-based model, i.e., \textit{SSVM}~\cite{junior2016specialized}, proposes to control the empirical risk to ensure a finite risk of the unknown classes. 
On the other hand, \textit{OSNN}~\cite{junior2017nearest} extends the nearest-neighbor classifier to the OSR scenario by fully incorporating the ability to recognize samples belonging to classes that are unknown during training. A generalized sparse-representation based classification algorithm is then proposed to make use of the generalized Pareto extreme value distribution for OSR.
The extreme value theory-based models are a special category derived from the statistical modeling methods for statistical extremes~\cite{de2007extreme}. Among them, EVM~\cite{rudd2018extreme} is the most representative one, which utilizes data-distribution information with a strong theoretical foundation.

\subsubsection{DNNs-based Methods}
With the popularity of deep learning techniques, the deep neural networks have became a powerful alternative to implement OSR systems. Among them, the \textit{OpenMax}~\cite{bendale2016towards} is one of the pioneers that consider estimating the probability of an input sample being from an unknown class by using deep neural networks. 
Recently, some NN-based methods also integrate the generative models, e.g., generative adversarial networks (GANs) or Variational autoencoders (VAEs), into OSR models to construct more robust classifiers. For example, the \textit{G-OpenMax}~\cite{ge2017generative} extends the \textit{OpenMax} by synthesizing novel unknown samples from the known class data with GANs, such that the classifier can be endowed with some priori knowledge of the distribution or representation of unknown classes. 
Slightly different from the \textit{G-OpenMax}, Neal \textit{et al.}~\cite{neal2018open} propose to synthesize samples that are close to the training set yet not belonging to any training class to enrich the decision ability of classifier. 
Besides, the \textit{OLTR}~\cite{liu2019large} focuses a long-tail and open-ended distribution and proposes a novel model capable of handling various tasks such as OSR, imbalanced classification, and few-shot learning in one integrated algorithm. 
Hassen \textit{et al.}~\cite{hassen2018learninga} aim at learning more powerful representation of samples by utilizing class centers to obtain the features of known classes for OSR. 
%
 
\section{Proposed Method}
\label{sec:proposed_method}
We first give the problem formulation of MOSR task. Then, we introduce our proposed method in detail. Concretely, it consists of two parallel encoders dealing with numeric features and textual features of malware samples, respectively. The encoded features of such two modalities are further fused to be an enhanced feature representation that is endowed with more diversities. Moreover, the contrastive sampling and $\rho$-bounded enclosing sphere regularizations are applied to modify the obtain representation into one primary space and an associated sub-space, i.e., discriminative and exclusive spaces, to facilitate the classification and detection tasks, respectively. 
Last, a modified distance-based detection mechanism is further introduced.

\begin{figure}[t]
  \centering
  

\subfigure[Digit 1]{
   \label{fig:subfig:a} 
   \includegraphics[width=1.0in]{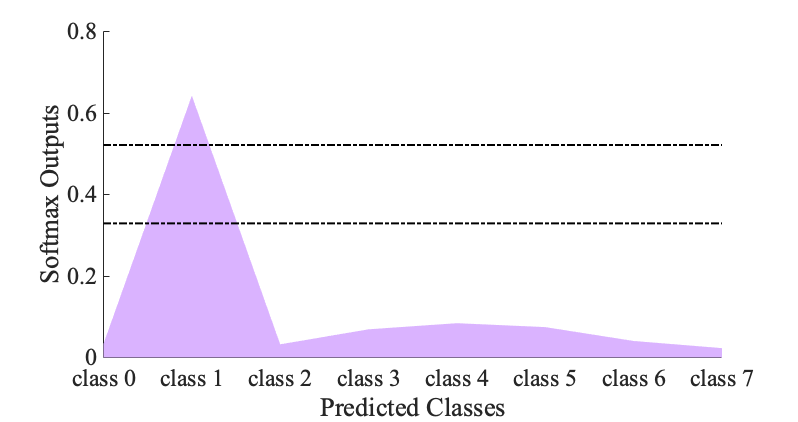}}
\subfigure[Digit 2]{
   \label{fig:subfig:b} 
   \includegraphics[width=1.0in]{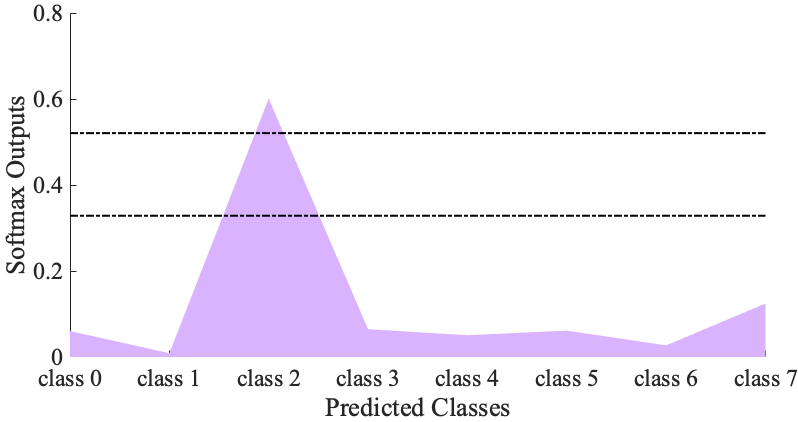}} 
\subfigure[Digit 3]{
   \label{fig:subfig:c} 
   \includegraphics[width=1.0in]{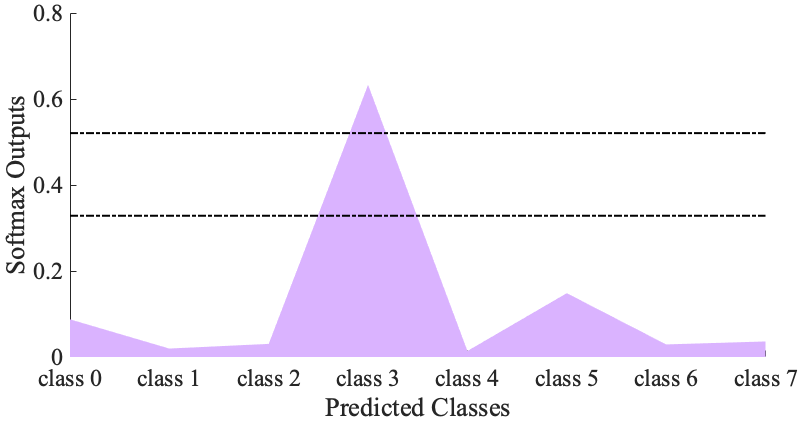}}

\subfigure[Digit 4]{
   \label{fig:subfig:d} 
   \includegraphics[width=1.0in]{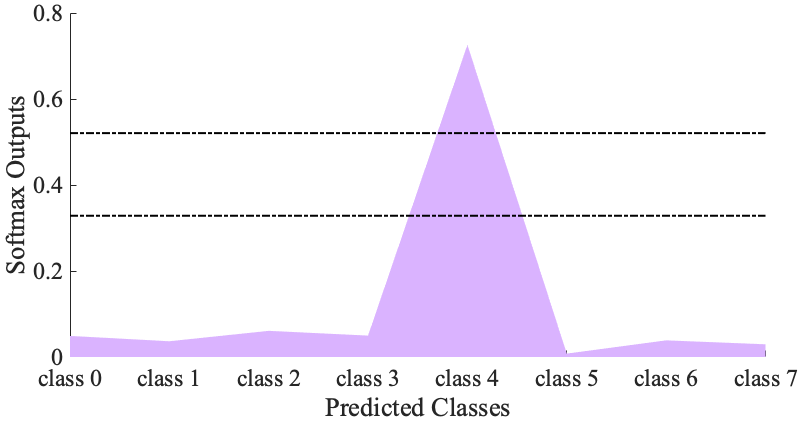}} 
\subfigure[Digit 5]{
   \label{fig:subfig:e} 
   \includegraphics[width=1.0in]{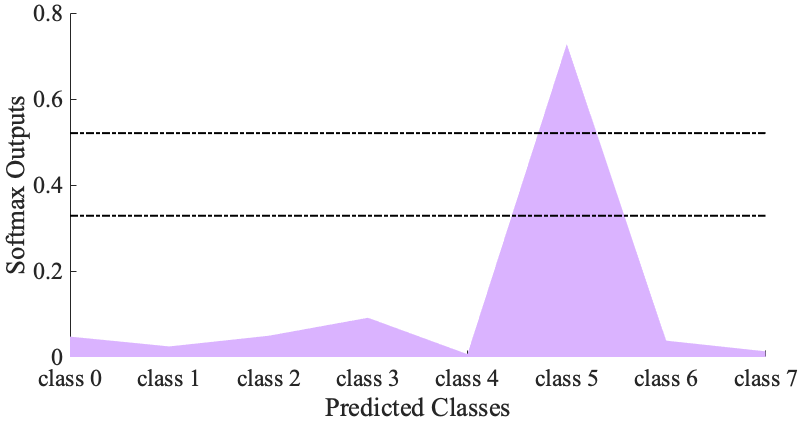}}
\subfigure[Digit 6]{
   \label{fig:subfig:f} 
   \includegraphics[width=1.0in]{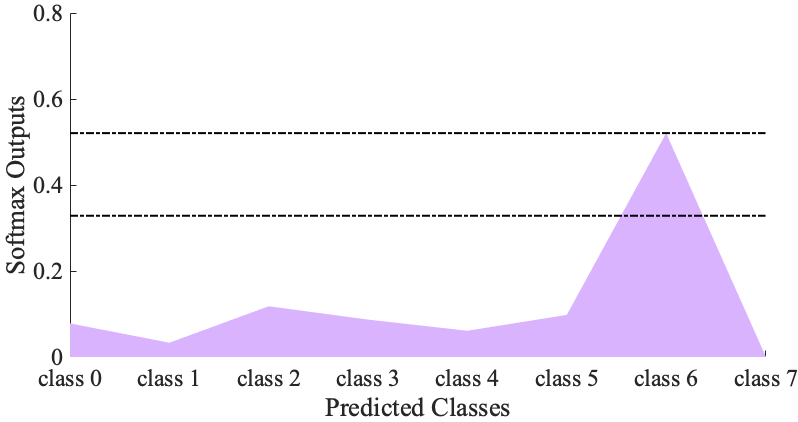}} 

\subfigure[Digit 7]{
   \label{fig:subfig:g} 
   \includegraphics[width=1.0in]{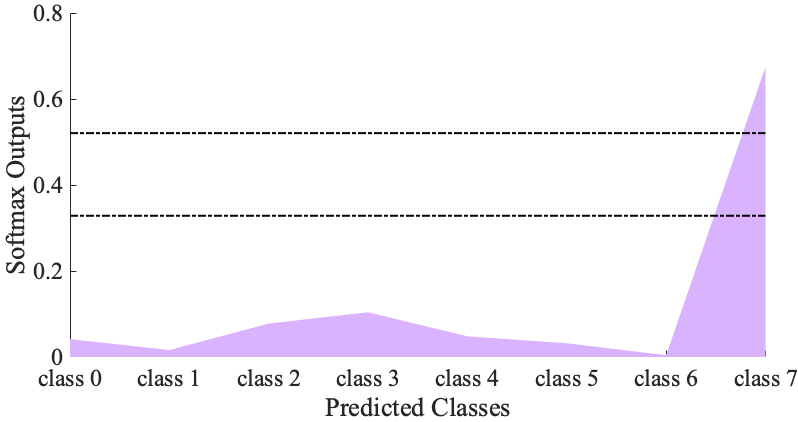}}
 \subfigure[Digit 8 (unknown)]{
   \label{fig:subfig:h} 
   \includegraphics[width=1.0in]{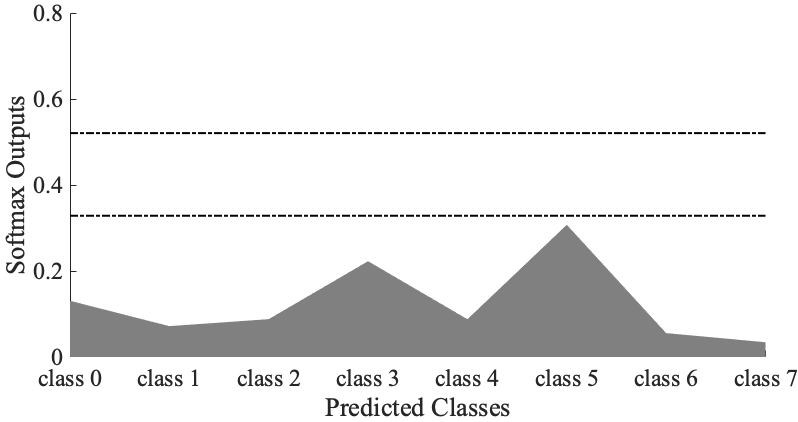}} 
\subfigure[Digit 9 (unknown)]{
   \label{fig:subfig:i} 
   \includegraphics[width=1.0in]{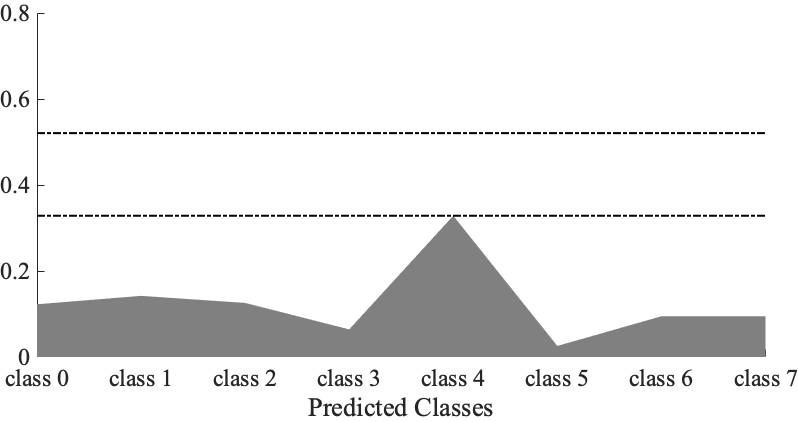}}

\caption{Visualization of softmax outputs of digits on known classes `1'$\sim$`7' and unknown classes `8' and `9'. 
(a)$\sim$(g) marked in violet are known classes, and (h), (i) marked in gray are unknown classes (better viewed in color).}
  \label{mnist_prob}
\end{figure}

\subsection{Problem Formulation}
Given a malware dataset $\mathcal{D}_{tr}$ consisting of $n$ malware samples labeled from $K$ known families $\mathcal{D}_{tr} = {\{ ( x_i ,  l_i ) \}}_{i=1}^{n}$, where $x_i$ is a malware sample of family $l_i$, and $l_i \in F_k = \{ f_1, f_2, \ldots, f_K \}$. 
Specifically, $x_i$ usually contains a set of extracted features in numeric modality. Hereby, we further consider another modality of malware features from tokenized behaviors and dynamics to enhance the diversity of malware feature space. 
As a common practice, conventional malware classification approaches usually rely on a well-trained classifier $\mathcal{P}(\cdot; \cdot)$, that outputs probability distribution on $F_k$ when given a malware sample $x$ of known families. Then, it can calculate $\hat{l}$, the predicted family of $x$ using the obtained probability distribution. 
Specifically, $\mathcal{P}(\cdot; \cdot)$ can be trained on $\mathcal{D}_{tr}$ with the cross-entropy loss objecting to:
\begin{equation}
\mathop{\min}\limits_{\Theta} \; \frac{1}{n} \cdot \sum_{i=1}^{n} \sum_{k=1}^{K} - l_{ik} \log {\mathcal{P}(x_i;\Theta)}_{k}, \label{eq: classifier loss in pf} 
\end{equation}
where $l_{ik}$ corresponds to the $k$-th index/value of the normalized one-hot vector of $l_i$, and $\Theta$ denotes the trainable parameters of $\mathcal{P}(\cdot; \cdot)$. 
In the open-set scenario, since the real family of input malware sample $x$ may not be known ones, i.e., $l \notin F_k$, a detection process is also needed to distinguish whether $x$ comes from known families or falls into a novel unknown family at the same time. 
%
%
%
Practically, a probability-based detection approach borrowed from computer vision domain is usually adopted by conventional methods, to detect whether $x$ belongs to novel unknown families, i.e., by applying a detector $\mathcal{DET}(\cdot)$ on the probabilistic outputs of the classifier:
\begin{equation}
\mathcal{DET}(\mathcal{P}(x)) \le \delta,
\end{equation}
where $\delta$ is a pre-defined threshold that determines the predicted malware sample $x$ to be unknown families if the condition satisfies or known families otherwise. 

Such a probability-based detection is based on an assumption that the margins or differences are large enough between known and unknown families, such that the predicted probability of $x$ belonging to known families may vary significantly if $x$ comes from a novel unknown family, i.e., $l \notin F_k$. This assumption is strongly hold in computer vision domain and makes the novelty detection robust. As demonstrated in Fig.~\ref{mnist_prob}, we use the most simple image dataset MNIST~\cite{lecun1998mnist} as an example, from which digits `1' $\sim$ `7' are set as known classes, and digits `8' and `9' are set as unknown classes. We visualize the softmax outputs of testing digitals of each classes by a toy neural network, i.e., can be approximated as the probability distribution, and we can observe a high peak exists in the predicted probability distribution of known classes (Fig.~\ref{fig:subfig:a}$\sim$~\ref{fig:subfig:g}), while the predicted probability distribution of unknown classes is much flatter and the peak is obviously low (Fig.~\ref{fig:subfig:h}$\sim$~\ref{fig:subfig:i}). 
However, the performance may degrade dramatically in malware recognition task due to the limited divergence among different malware, even between known and unknown families. In our method, we resort to constructing more powerful representation of malware samples which is more representative and discriminative for down-stream MOSR task. Moreover, a more efficient distance-based detection mechanism is further introduced to mitigate the detection degradation.

\begin{figure*}[t]
	\centering
	\includegraphics[width=0.87\textwidth]{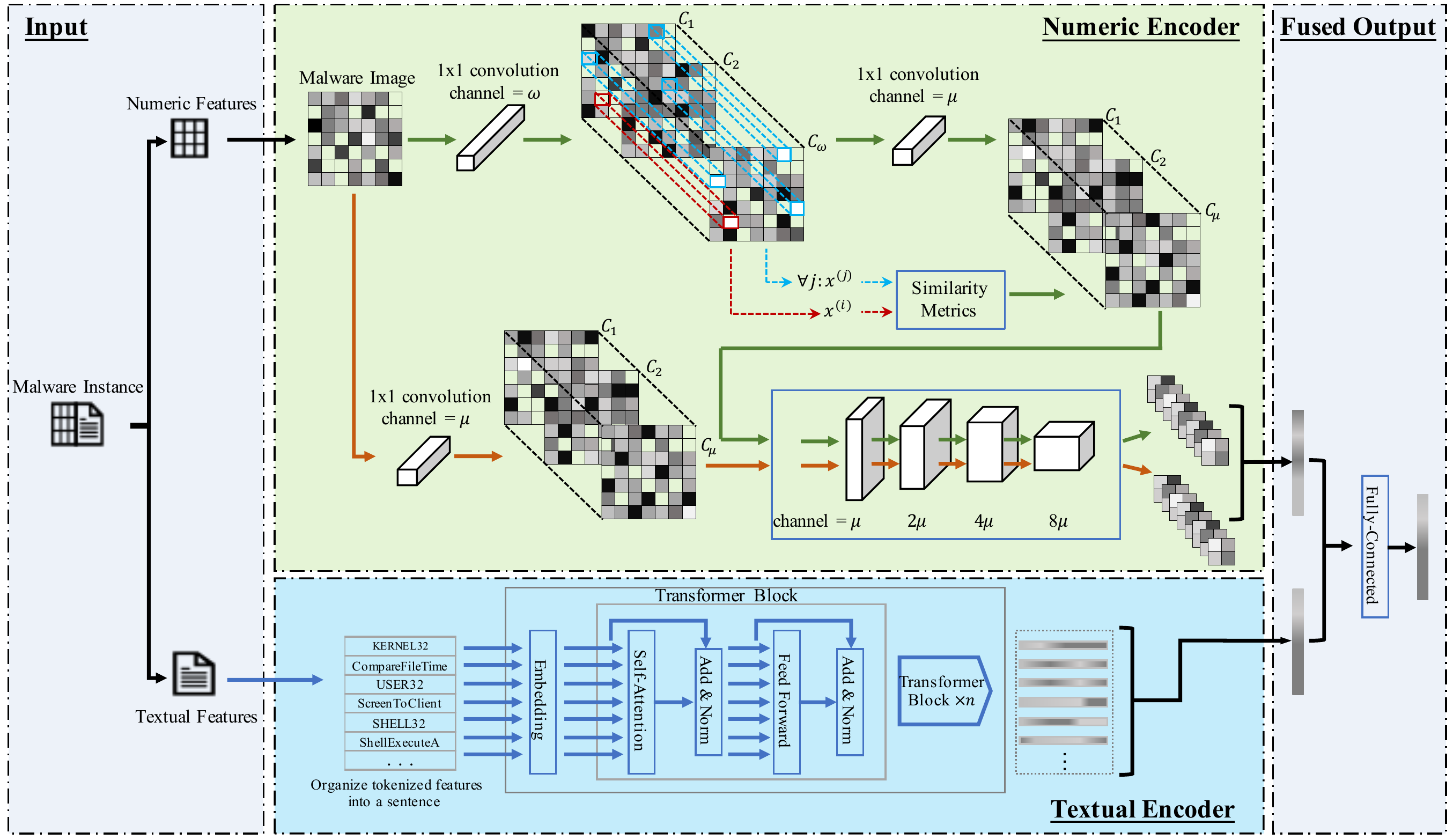}
	\caption{Multi-modal encoders architecture: a re-designed CNN-based numeric encoder (upper part) and a language model-based textual encoder (lower part) are parallelly placed to deal with extracted malware numeric characteristics and tokenized textual behaviors/dynamics, respectively. Within the numeric encoder, two branches of the re-designed global receptive module (green arrows) and local receptive module (orange arrows) jointly extract multi-correlations of organized malware images with nearby and long-distance dependence. Within the textual encoder, the organized malware sentences are represented into high-level and self-attended semantic vectors. Such two representations are further fused for the down-stream recognition (better viewed in color).}
	\label{fig: multimodal encoder}
\end{figure*}

\subsection{Multi-modal Encoders Architecture}
Conventional methods fail in MOSR task especially in detection because the divergence of malware families is usually very small compared to other domains. We argue such a limited divergence comes from two facts. First, the attack techniques of most malware are quite similar, which makes the malware features highly homogeneous. 
Second and more easily overlooked, existing methods only utilize the extracted numeric features, e.g., statical characteristics, to construct the feature space and train an off-the-shelf classifier based on it. 
Such a common practice is not robust and neglect constraining necessary properties on the classifier to fit the special requirement of malware recognition. 
%
In our method, we consider fully making use of comprehensive malware features in both extracted numeric characteristics and tokenized textual behaviors/dynamics to better represent the malware feature space. Specifically, as shown in Fig.~\ref{fig: multimodal encoder}, a multi-modal encoders architecture consisting of a re-designed CNN-based numeric encoder, and a language model-based (e.g., BERT~\cite{devlin2018bert}) textual encoder, are parallelly placed to better extract and fuse malware features in both modalities.

\subsubsection{Numeric Encoder}
Existing MFR methods usually re-organize extracted numeric features as malware images, and make use of CNN backbones to explore the statistical and spatial correlations of these features within each malware sample. In our numeric encoder, we also follow this fashion. 
%
Different from existing methods that use an off-the-shelf CNN architecture as the feature extractor, we re-design a more adaptive architecture with multi-scale properties to better extract and represent malware samples. Concretely, two additional necessary properties are claimed in our numeric encoder for malware domain: \textit{long-distance dependence} and \textit{non-downsampling}. 

First, conventional CNN architectures are intrinsically tuned for capturing local dependences, e.g., a pixel and its nearby pixels in an image, and thus lack the ability for efficiently capturing long-distance dependence among remote features of a sample. However, different from the computer vision domain, the numeric features of malware samples are usually extracted discretely and may not have a strong locality similar to real images. Thus, the malware samples are tend to be posed with multi-scale correlations among features. 
%
To deal with this challenge, we re-design the CNN architecture with both global and local receptive fields in a multi-scale manner. As shown in Fig.~\ref{fig: multimodal encoder}, the pipeline marked with green arrows is endowed with the global receptive capacity inspired by the classic non-local means~\cite{buades2005non}. Specifically, the operation within the global receptive module can be described as:
\begin{equation}
\tilde{x_{i}} = \frac{1}{\mathcal{S}(x)} \cdot \sum_{\forall j}^{} f(\mathcal{T_{\omega}}(x)_{i}, \mathcal{T_{\omega}}(x)_{j}) \cdot \mathcal{T_{\mu}}(x)_{j},
\end{equation}
where $x$ is an input malware image, i.e., in grey-scale form with height $\mathcal{H}$ and width $\mathcal{W}$. $\mathcal{T_{\omega}}(\cdot)$ and $\mathcal{T_{\mu}}(\cdot)$ are two functions that transform $x$ into two feature maps with $\omega$ and $\mu$ channels, respectively. $i$ is the feature index of our target position in a feature map and $j$ enumerates all possible positions with both nearby and long-distance features. $f(\cdot, \cdot)$ calculates the similarity of each $i/j$ pair and $\mathcal{S}(x)$ is a normalization factor. In our method, we practically implement $\mathcal{T_{\omega}}(\cdot)$ and $\mathcal{T_{\mu}}(\cdot)$ by trainable $1 \times 1$~convolutions, and choose the Gaussian function as the similarity metric, i.e., 
\begin{equation}
f(\mathcal{T_{\omega}}(x)_{i}, \mathcal{T_{\omega}}(x)_{j}) = e^{(T_{\omega}(x)_{i})^{^\mathsf{T}} T_{\omega}(x)_{j}}, 
\end{equation}
hereby, the normalization factor $\mathcal{S}(x)$ can be explicitly set as the accumulation of similarities, i.e., $\sum_{\forall j}^{} e^{(T_{\omega}(x)_{i})^{^\mathsf{T}} T_{\omega}(x)_{j}}$. The global receptive module is fairly intuitive compared to classic kernel-based convolution operations, whose receptive field is typically bounded within a local region, i.e., $i - \kappa \le j \le i + \kappa$, where $\kappa$ denotes the kernel size, while the global receptive module enumerates every $j \in \left \{ 1, 2, \dots , \mathcal{H}\ast \mathcal{W} \right \} $ to capture the long-distance dependences of malware features. 

Second, our local receptive module that marked with orange arrows is similar with conventional CNN architectur. It integrates a sequence of convolutional blocks to capture the local dependences of malware features, of which each block is a stacked structure of three layers, namely, 2-dimensional convolutional layer, batch normalization (BN) layer~\cite{DBLP:conf/icml/IoffeS15}, and ReLU activation layer~\cite{DBLP:conf/icml/NairH10}. 
Differently, we do not downsample the generated malware feature maps with pooling layers, which are widely used in conventional settings. This is because the organized malware image is usually extracted discretely, i.e., the numeric features are statistical results of bytes or instruction calls, thus the the pooling operation may filter useful and important information out. 
Instead, we increase the strides of certain convolutional layers, e.g., set as 2, to generate a similar output as downsampling. Moreover, the kernel size of each convolutional layer are increased synchronously, i.e., 5$\times$5, to achieve a larger local receptive field. 

In summary, the encoding process of the numeric encoder can be conceptually described as:
\begin{equation}
z^{num} = \operatorname{ReLU} ( W_{ln}^{^\mathsf{T}} \ast \left [ \mathcal{GM}(x), \mathcal{LM}(x) \right ]),
\end{equation}
where $z^{num}$ denotes the encoded representation of sample~$x$. $\mathcal{GM}(\cdot)$ and $\mathcal{LM}(\cdot)$ are the global and local receptive modules, respectively. $\left [ \cdot, \cdot \right ]$ denotes the concatenation operation and $W_{ln}^{^\mathsf{T}}$ is a linear transformation that projects the concatenated feature map to a vector.

\subsubsection{Textual Encoder}
Different from existing MFR methods, we further organize tokenized textual features of each malware sample to a sentence, within which the words explicitly represent some import domains of malware behaviors and dynamics including libraries, modules, functions, etc., that are \textit{sequentially} and \textit{contextually} used. It should be noted that, this assumption of sequentialization and contextualization relies on the co-dependencies of malware behaviors and dynamics triggered by the inherent attack mechanism of different families. 

In our method, we make use of organized malware sentences to capture such co-dependencies concerned with different malware families, and thus to enhance the diversity of malware feature space. Our textual encoder borrows from \textit{BERT}~\cite{devlin2018bert} architecture and is composed of multiple Transformer Encoder blocks~\cite{DBLP:conf/nips/VaswaniSPUJGKP17} serially. Specifically, each block contains a self-attention layer and a feed-forward layer, that are wrapped by residual connections with layer normalization~\cite{DBLP:journals/corr/BaKH16}, respectively, i.e., $ \operatorname{LayerNorm} ( s + \operatorname{Layer} ( s ))$, where $s$ is the input malware sentence. 
For each given word embedding vector $s_i$ in a sentence, the self-attention layer computes a set of $\left \{ \textit{query}: q_i = Q s_i, \textit{key}: k_i = K s_i, \textit{value}: v_i = V s_i \right \}$, where $Q$, $K$, and $V$ are learnable matrices. 
Then, it computes weights using the softmax of the correlation between $k_i$ and each \textit{key} vector to produce $\tilde{v}_{i}$, i.e., a weighted sum of value vectors as: 
\begin{equation}
	\tilde{v}_{i} = \sum_{j=1}^{n} \frac{\exp ( q_i k_j )}{\sum_{m=1}^{n} \exp ( q_i k_m )} v_j.
\end{equation}
The feed-forward layer consists of two fully-connected layers with a ReLU activation in between is further placed, i.e.: 
\begin{equation}
	\operatorname{FCN} ( v ) = \max (\vec{0} , v W_1 + b_1 ) W_2 + b_2 , 
\end{equation}
where $W_1$, $W_2$ are weight matrices, and $b_1$, $b_2$ are biases. By stacking multiple blocks serially, we can finally obtain the self-attended encoding of each malware sentence with the textual encoder, conceptually denoted as:
\begin{equation}
z^{tex} = \operatorname{ReLU} ( W_{lt}^{^\mathsf{T}} \ast \mathcal{TM}(s)),	
\end{equation}
where $\mathcal{TM}(\cdot)$ denotes the stacked blocks and $W_{lt}^{^\mathsf{T}}$ is a linear transformation that projects the array of word embeddings of malware sentence to a vector. 

The encoded representations of the malware image and malware sentence, i.e., $z^{num}$ and $z^{tex}$, are further concentrated and fused by fully-connected layers to generate a unified malware encoding $z$ as:
\begin{equation}
\begin{aligned}
z = \operatorname{FCN} \left.([ 
    \operatorname{ReLU} ( W_{ln}^{^\mathsf{T}} \ast \left [ \mathcal{GM}(x), \mathcal{LM}(x) \right ]), \right.\\
      \left. \cdots \operatorname{ReLU} ( W_{lt}^{^\mathsf{T}} \ast \mathcal{TM}(s)) \right.]),
\end{aligned}
\end{equation}
which will be used for down-stream classification and detection tasks in our proposed MOSR method.

\subsection{Dual-embedding Space Learning}
Given the multi-modal inputs for malware families, i.e., ${\{ ( x_i, s_i, l_i ) \}}_{i=1}^{n}$, where $x_i$ and $s_i$ denote the organized malware image and sentence, respectively. $l_i$ is the family label and $l_i \in F_k = \{ f_1, f_2, \ldots, f_K \}$. 
To handle the classification task, the multi-modal encoders, i.e., denoted as $\operatorname{E}(\cdot, \cdot)$ for simplicity, is first used to embed $x_i$ and $s_i$ into a common vector space as $z_i$, where a classifier $\operatorname{C}(\cdot)$, i.e., an simple sequence of multiple fully-connected layers, further maps those embedded vectors to their corresponding families. Such classification process can be trained by the following cross-entropy loss as:
\begin{equation}
\operatorname{Loss}_{Cls}(\Theta_{\operatorname{E}}, \Theta_{\operatorname{C}}) = \frac{1}{n} \cdot \sum_{i=1}^{n} \sum_{k=1}^{K} - p_{k}(l_i) \log q_{k}(x_i, s_i), 	
\end{equation}
where $\Theta_{\operatorname{E}}$ and $\Theta_{\operatorname{C}}$ are trainable parameter-set of $\operatorname{E}(\cdot, \cdot)$ and $\operatorname{C}(\cdot)$, respectively. 
%
$p_{k}(\cdot)$ and $q_{k}(\cdot)$ are defined as:
\begin{align}
	& p_{k}(l_i)  = \begin{cases}  
	1,          & l_i = k          \\
	0,          & l_i \neq k       
	\end{cases}, \\
	& q_{k}(x_i, s_i)  = {\operatorname{C}(\operatorname{E}(x_i, s_i))}_k, 
\end{align}
which indicates whether the label $k$ is $l_i$, and the probability of label $k$ predicted by the classifier, respectively. 
To improve the classification performance and enable our model to accurately detect novel unknown malware families, we further construct one primary space and an associated sub-space, namely, \textit{discriminative} and \textit{exclusive} spaces, by using contrastive sampling and $\rho$-bounded enclosing sphere regularizations. 


\subsubsection{Discriminative Primary Space}
As mentioned in Sect.~\ref{sec:introduction}, another key challenge in MOSR lies in the limited divergence of malware features, even between known and novel unknown families. 
In our method, to further improve the model discrimination ability for malware samples between different families, we propose to integrate contrastive sampling to enlarge the divergence of malware feature space, i.e., learning a discriminative space, especially for the task of classification. 
Specifically, in the training phase, we randomly select two sample pairs $\left \{ (x_i, s_i), (x_{i}^{+}, s_{i}^{+}) \right \}$ and $\left \{ (x_i, s_i), (x_{i}^{-}, s_{i}^{-}) \right \}$ for each sample $(x_i, s_i)$, where their families satisfy $l_i^{+}=l_i$ and $l_i^{-} \neq l_i$. It is noticed that, this sampling is not needed in the testing phase. 
To construct the discriminative space, we first embed the triplet into the common feature space by utilizing the multi-modal encoders $\operatorname{E}(\cdot, \cdot)$ as:
\begin{equation}
\begin{cases}
	z_i   & = \operatorname{E}(x_i, s_i),    \\
	z_i^+ & = \operatorname{E}(x_i^+, s_i^+),  \\
	z_i^- & = \operatorname{E}(x_i^-, s_i^-). 
\end{cases}
\end{equation}
Next, it is expected that within the discriminative space, the model learns more discriminative representations such that similar samples stay close to each other, while dissimilar ones are far apart. For this purpose, we train the features with the following constrains as:
\begin{align}
& \mathop{\min} \limits_{\Theta_{\operatorname{E}}} \| z_i - z_i^+ \|_2 ,  \\ 
& \mathop{\max} \limits_{\Theta_{\operatorname{E}}} \| z_i - z_i^- \|_2 , 
\end{align}
where $\| \cdot \|_2$ denotes the $L_2$-distance as the similarity metric. Eq.~(14) minimizes the similarity of feature pair $z_i$ and $z_i^+$, which can force them to be close to each other. Eq.~(15) maximizes the similarity of feature pair $z_i$ and $z_i^-$, which can force them to be apart away from each other. We add the discriminative property into our training loss function as:
\begin{equation}
\operatorname{Loss}_{Disc}(\Theta_{\operatorname{E}}) = \sum_{i=1}^{n} \left( \| z_i - z_i^+ \|_2 - \| z_i - z_i^- \|_2 \right).
\end{equation}
With the gradient decreases of the loss function, $\| z_i - z_i^+ \|_2$ is trained to decrease while $\| z_i - z_i^- \|_2$ is trained to increases which can achieve the goal of discriminative representations.

\begin{figure}[t]
	\centering
	\includegraphics[width=0.47\textwidth]{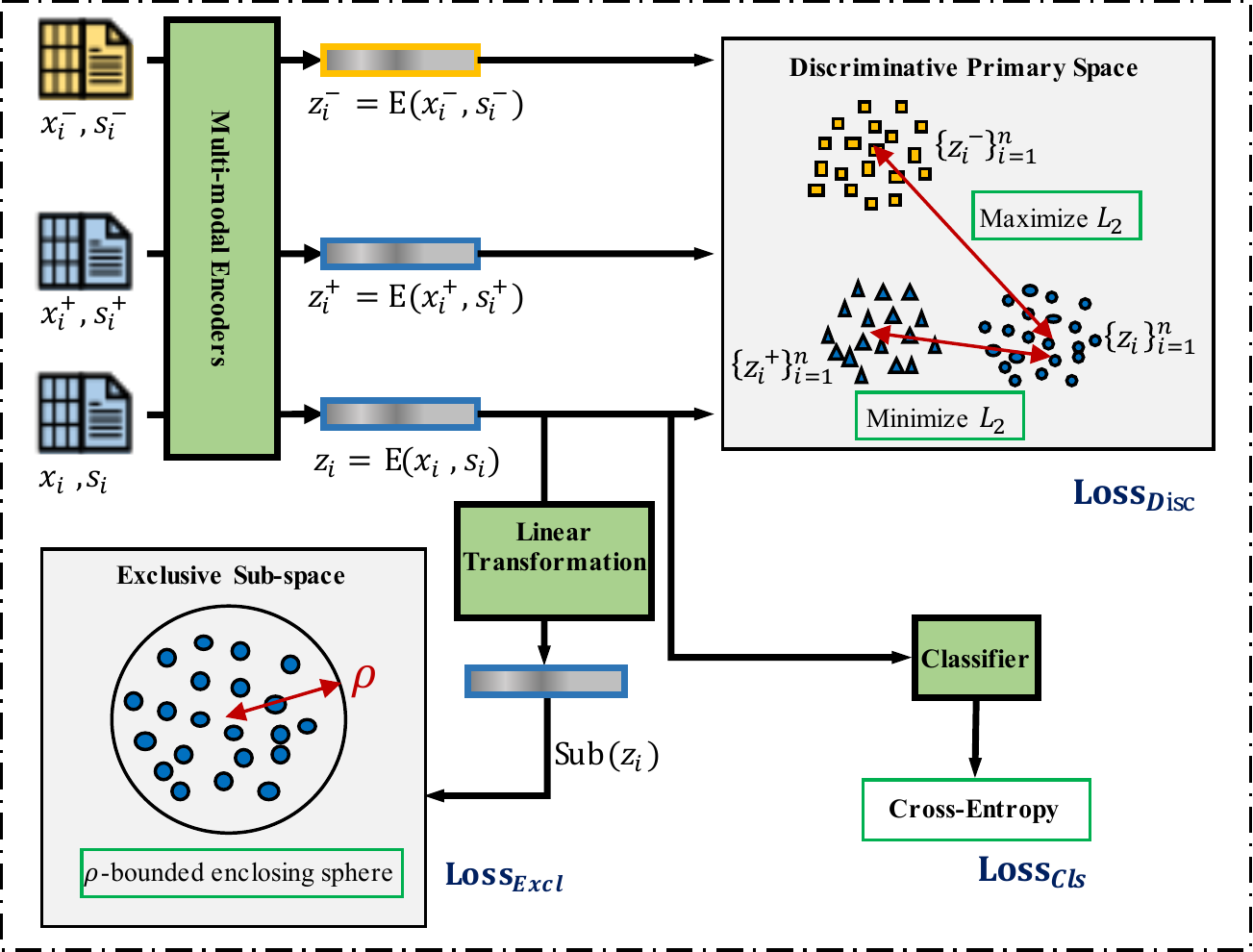}
	\caption{The training framework of our dual-embedding space involves the discriminative primary space, and an exclusive sub-space associated with it (better viewed in color).} 
	\label{fig: train}
\end{figure}

\subsubsection{Exclusive Sub-space}
The above discriminative primary space helps deriving compact malware feature representations, i.e., ${\{ z_i \}}_{i=1}^{n}$, which are endowed with more diversities. Such properties are extremely useful for classification, while the model still lacks the ability to exclude novel unknown malware families that haven't been encountered during training. Inspired by the minimal sphere~\cite{van2021charge,wang2019coresets} that limits data inside a certain sphere, we propose to make use of the sample clusters of known families to improve the model detection ability for samples of unknown families. 
Specifically, we assume the known malware families can be bounded within a minimal enclosing sphere, and thus any sample with its features that are not covered by the sphere is considered from a novel unknown family. For this purpose, we train ${\{ z_i \}}_{i=1}^{n}$ with the following constrains as:
\begin{align}
\mathop{\max} \limits_{\Theta_{\operatorname{Sub}}, \rho} & \; |\{ \| \operatorname{Sub}(z_i) \|_2 \leq \rho \}| , \label{rho constraint} \\
\mathop{\min} \limits_{\Theta_{\operatorname{Sub}}, \rho} & \; \rho ,  	
\end{align}
where $\operatorname{Sub}(\cdot)$ is a linear transformation that embeds $z_i$ into an associated sub-space, namely, exclusive sub-space. $\| \cdot \|_2$ denotes the $L_2$-distance and $|\cdot|$ is the number of samples. The radius $\rho$ defines a sphere region where all $\operatorname{Sub}(z_i)$s from known families reside in. 
The Eq.~(17) maximizes the number of known malware samples within the sphere, and Eq.~(18) minimizes the radius $\rho$ to explore the feature boundary between known and unknown families. We add the above exclusive property into our training loss function as:
\begin{equation}
\begin{aligned}
	\operatorname{Loss}_{Excl}(\Theta_{\operatorname{Sub}}) = \sum_{i=1}^{n} \mathop{\max} \{ \| \operatorname{Sub}(z_i) \|_2 &-\rho , 0 \} + \rho \\
	&- \lambda \| \Theta_{\operatorname{Sub}} \|_2 .                                                
\end{aligned}
\end{equation}
In this loss, term $\mathop{\max} \{ \| \operatorname{Sub}(z_i) \|_2 -\rho , 0 \}$ indicates that if $\operatorname{Sub}(z_i)$ is not covered by the $\rho$-bounded sphere, the $\| \operatorname{Sub}(z_i) \|_2$ will be trained to decrease and thus push $\operatorname{Sub}(z_i)$ into the sphere; otherwise, i.e., if $\operatorname{Sub}(z_i)$ is covered by the sphere, it will not be trained. The radius $\rho$ is trained to minimized with decreasing loss. 
Moreover, a positive hyper-parameter $\lambda$ is used to prevent $\| \Theta_{\operatorname{Sub}} \|_2$ from becoming zero during training, because an extremely small $\| \Theta_{\operatorname{Sub}} \|_2$ can force all $z_i$s satisfy constraint of Eq.~(17) at a very starting point, which is not encouraged during training.

\subsubsection{Loss Function}
The overall loss function of our method integrates Eqs.~(10), (16), and (19) into a joint training framework (Fig.~4) as:
\begin{equation}
\begin{aligned}
	\operatorname{Loss} (\Theta_{\operatorname{E}}, \Theta_{\operatorname{C}}, \Theta_{\operatorname{Sub}}) & = \alpha \cdot \operatorname{Loss}_{Cls}(\Theta_{\operatorname{E}}, \Theta_{\operatorname{C}})  \\
	& + \beta \cdot \operatorname{Loss}_{Disc}(\Theta_{\operatorname{E}})              \\
	& + (1-\alpha -\beta) \cdot \operatorname{Loss}_{Excl}(\Theta_{\operatorname{Sub}}) , 
\end{aligned}
\end{equation}
where $\alpha$, $\beta$, and $1-\alpha -\beta$ are three positive hyper-parameters that control the balance of each sub-loss function, and we limit their summation as 1 to control the scale of them. The Adam optimizer~\cite{kingma2014adam} is used for the model training phase with the above loss function. Notably, the dual-embedding spaces involving $\operatorname{Loss}_{Disc}$ and $\operatorname{Loss}_{Excl}$ are only processed in the training phase for regularizing a more representative $\operatorname{E}(\cdot, \cdot)$, which resorts to a universal embedding for both classification and detection for our MOSR task.

\subsection{Distance-based Detection for Unknown Families}
\label{Distance-based Detection for Unknown Families}
The MOSR requires the model to first correctly detect given malware samples as either known or unknown families, and simultaneously classify those samples of known families into their corresponding families as accurately as possible. 
Based on our model, we first detect whether a malware sample belongs to known or unknown families. If the sample is identified as known families, the result of the classifier can be regarded as its corresponding family. 
For detection, we alternatively apply a distance-based mechanism rather than the widely used probability-based approach (it compares the family probability distribution of the classifier),  to better identify the novelty of unknown malware families.
Specifically, we first find the feature centroid of each malware family in the train-set $\mathcal{D}_{tr}$. Given a training malware sample~$(x, s, l) \in \mathcal{D}_{tr}$, we use its embedded feature~$\operatorname{E}(x, s)$ to calculate the feature centroids of the known families as:
\begin{align}
c_k = \frac{1}{N_k} \cdot \sum_{\mathcal{D}_{tr}, l = f_k} \operatorname{E}(x, s), 
\end{align}
where $N_k$ is the number of samples belonging to family $f_k$. 
Next, the multi-class classifier $\operatorname{C}(\cdot)$ is utilized to tentatively predict the belonging known family of sample $x$. 
The distance within the embedded feature space between $x$ and the feature centroid of the tentatively-predicted family is used by a detector $\mathcal{DET}(\cdot)$ to determine whether or not $x$ belongs to unknown families, as the following formula: \begin{align}
	\mathcal{DET}(x) = \begin{cases}
	\text{Known,}   & \| \operatorname{E}(x, s)-c_{\max \{ \operatorname{C}(\operatorname{E}(x, s)) \}} \|_2 \leq \delta \\
	\text{Unknown,} & \| \operatorname{E}(x, s)-c_{\max \{ \operatorname{C}(\operatorname{E}(x, s)) \}} \|_2 > \delta    
	\end{cases},
\end{align}
where $c_{\max \{ \operatorname{C}(\operatorname{E}(x, s)) \}}$ corresponds the feature centroid of the known family that determined by the maximum index/value of the predicted probability distribution, e.g., softmax outputs. $\delta$ is the threshold, which can be empirically set as the maximum distance between the feature centroid and all samples of the tentatively-predicted family, i.e., 
\begin{align}
	\delta = \max_{\mathcal{D}_{tr}} \| \operatorname{E}(x, s)-c_l \|_2. 
\end{align}

It should be noted that, such a distance-based detection mechanism can well mitigate the detection degradation that relies on the conventional probability-based approach, of which the classifier usually tends to produce overly high probability distributions in malware recognition. 
The overall algorithm of the above MOSR process, i.e., classification and detection, is briefly demonstrated in Algorithm~\ref{alg: osr}.

\begin{algorithm}[tbh]
	\caption{Malware OSR}\label{alg: osr}
	\begin{algorithmic}[1]
		\REQUIRE Trained networks $\operatorname{E}(\cdot, \cdot)$ and $\operatorname{C}(\cdot)$ $\Leftarrow$ Eq.~(20).
		\REQUIRE Feature centroids $c_1$,\ldots,$c_k$,\ldots$c_K$ $\Leftarrow$ Eq.~(21).
		\REQUIRE Threshold $\delta$ $\Leftarrow$ Eq.~(23).
		\STATE $z = \operatorname{E}(x, s)$, 
		\STATE Find the tentatively-predicted family $k$ that satisfies 
		\begin{align*}
			{\operatorname{C}(z)}_k = \max {\{ {\operatorname{E}(z)}_i \}}_{i=1}^{K},
		\end{align*}
		\IF{$\| z - c_{k} \|_2 \leq \delta$}
		\STATE Identify $x$ as a sample of the known family $k$,
		\ELSE
		\STATE Identify $x$ as a sample of unknown families.
		\ENDIF
	\end{algorithmic}
\end{algorithm}

\section{Experiments}
\label{sec:experiments}

\subsection{Datasets and Settings}
Our method is evaluated on two malware datasets including \textit{Mailing}~\cite{nataraj2011malware}, and our proposed \textit{MAL-100$^{+}$}. The basic descriptions and experimental settings of these two dataset are listed as follows.

\subsubsection{\textit{Mailing Dataset}}
The \textit{Mailing} is a widely used malware dataset containing 9,339 malware samples from 25 families in total. It groups the malware binaries sequences with different length as 8-bit malware vectors and then further reshapes these vectors into grayscale images. 
To evaluate, the first 15 families containing 8,017 malware samples are chosen as the known families for training and testing of classification. Specifically, we randomly select 80\%, i.e., 6400 samples, from these known families to form the train-set, and the remaining 20\%, i.e., 1,617 samples, are used to form the test-set. 
The remaining 10 families covering 1,322 malware samples are chosen as the unknown families and are mixed with the above train-set for detection jointly. 
Notably, the original grayscale malware images are in different size due to the non-uniform binaries sequences. To unify the input scale, each malware image is further resized into 32$\times$32. For those evaluated competitors that do not require malware images as the input, we flatten them into 1-d vectors that with the same length and feed them properly. It is noticed that since the \textit{Mailing} contains no textual features, we thus only consider numeric features in the dual-embedding of our method.

\subsubsection{\textit{MAL-100$^{+}$} Dataset}
In this paper, we enrich our previously proposed large-scale malware dataset, i.e., \textit{MAL-100}, with multi-modal characteristics and contribute an incremental version dubbed \textit{MAL-100$^{+}$}. Both datasets jointly fill the gap of lacking large-scale and multi-modal open-set benchmark in the malware recognition domain.
%
Concretely, it consists of a wide range of 100 malware families covering more than 50 thousand samples. To construct the dataset, the PE malware files from VirusSign \cite{VirusSign} are first collected. Next, the VirusTotal \cite{VirusTotal} and AVClass \cite{sebastian2016avclass} is applied to identify and label each collected PE file with a unique family group. Last, we make use of the widely used malware feature extraction tool, i.e., \textit{Ember}~\cite{anderson2018ember}, to extract static characteristics for each sample. In our dataset, we further organize the obtained malware characteristics into conventional numeric features to form the malware images, and tokenized textual features to form our newly defined malware sentences. Ideally, such two modalities of data should have the potential to be fused to enhance the malware feature space, of which each malware image is resized into 25$\times$25, and each malware sentence is represented by a fused 1-d word embedding with the dimensionality of 1,024. 

\begin{figure}[t]
  \centering

\centerline{  
\subfigure[Family distribution]{
   \label{fig:subfig:a} 
   \includegraphics[width=0.37\textwidth]{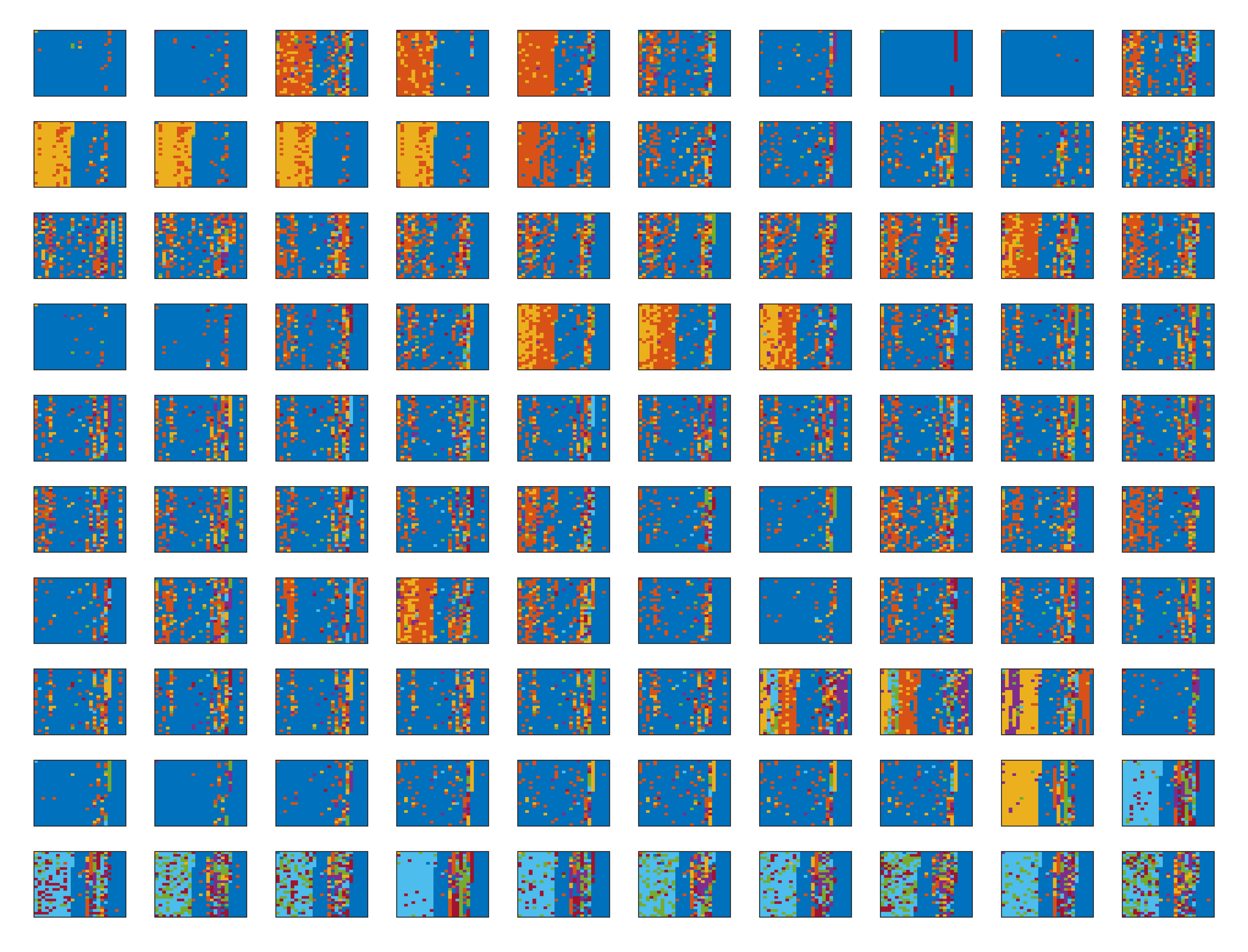} \label{fig: data_view1}}}\hspace{3mm}

\centerline{  
\subfigure[Sample distribution]{
   \label{fig:subfig:b} 
   \includegraphics[width=0.42\textwidth]{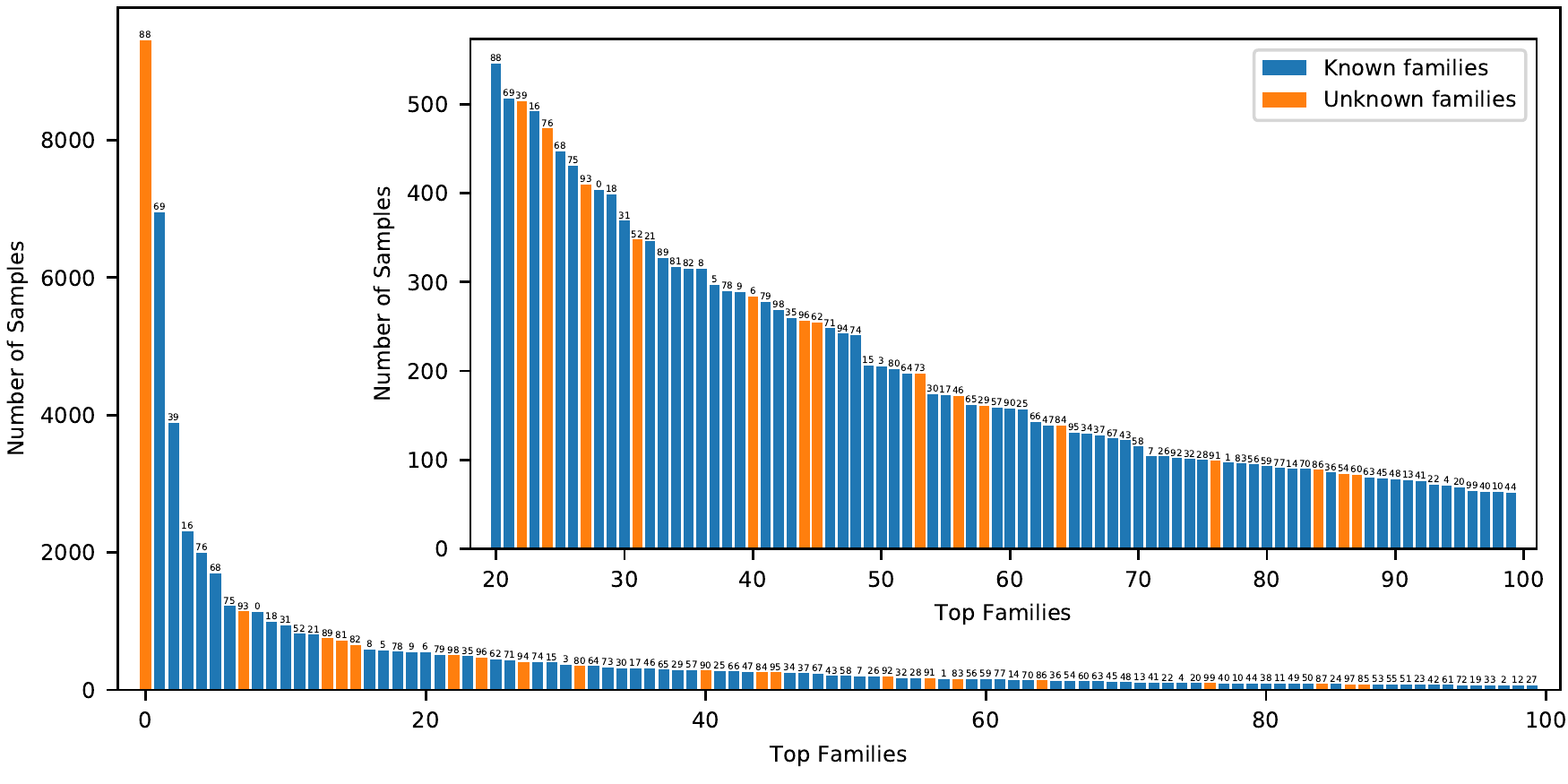}} \label{fig: data_view2}}
   
\caption{The visualization of the proposed malware dataset. Subfigure (a) denotes the family distribution in a 10$\times$10 grid, where each presented malware image corresponds to the mean-sample of each family. Subfigure (b) denotes the sample distribution in descending order, where the family labels are placed on the top of bars individually. We zoom-in the last 80 families for better demonstration (better viewed in color and zoom mode).}
  \label{fig: data_view}
\end{figure}

To better view the proposed dataset, we obtain the mean-sample of each malware family, i.e., by calculating the averaged malware image in a family-wise manner, and visualize the family distribution in as 100 grey-scale images in Fig.~\ref{fig: data_view}(a). We can observe that each family contains sufficient malware features and the variances are more apparent, which indicates a better generalization and representation of the proposed dataset. 
Moreover, we also visualize the sample distribution in Fig.~\ref{fig: data_view}(b), where we list the number of samples of each malware family in descending order bars. It can be observed that the proposed dataset follows a typical long-tail distribution similar to most real-world scenarios, which suggests our dataset provides a more practical and challenging setting. In summary, the proposed dataset is expected to contribute the future MOSR research with more extension ability.
To evaluate, the first 80 families (in ID orders) containing 39,346 malware samples are chosen as the known families for training and testing of classification. Among them, we randomly select 80\%, i.e., 31,523 samples, from these known families to form the train-set, and the remaining 20\%, i.e., 7,823 samples, are used to form the test-set. 
Similarly, the remaining 20 families covering 17,135 malware samples are chosen as the unknown families for detection.

\subsection{Evaluation Criteria}
The MOSR problem focuses on two goals: 1) Classify malware samples of known families as accurately as possible, which is referred as the family-wise classification task introduced in sect.~\ref{subsubsec:family-wise_classification_for_malware_recognition}; and 2) Detect malware samples of unknown families with high accuracy, which is referred as the unknown detection task introduced in sect.~\ref{subsec:Detection of Unknown Malware Families}

\subsubsection{Family-wise Classification}
The performance is evaluated by the classification accuracy $Cls_{Acc}$ defined as:
\begin{equation}
Cls_{Acc} = \frac{N_{correct}}{N_{sample}},
\end{equation}
where $N_{sample}$ denotes the total number of testing malware samples of known families and $N_{correct}$ is the number that are correctly classified. This definition corresponds to the conventional accuracy of most malware classification models.

\subsubsection{Unknown Detection}
To calculate the detection accuracy $Det_{Acc}$, we need to consider the testing malware samples of both known and unknown families. Specifically, we first obtain the true positive rate (TPR) for the samples of known families:
\begin{equation}
	TPR^{(k)} = \frac{TP^{(k)}}{TP^{(k)}+FN^{(k)}},
\end{equation}
where $TP^{(k)}$ and $FN^{(k)}$ are the true positive and the false negative for known families, respectively. 
Next, the true negative rate (TNR) for the samples of unknown families is obtained as:
\begin{equation}
	TNR^{(u)} = \frac{TN^{(u)}}{TN^{(u)} + FP^{(u)}},
\end{equation}
where $TN^{(u)}$ and $FP^{(u)}$ are the true negative and the false positive for unknown families, respectively. 
The detection accuracy $Det_{Acc}$ is then can be calculated as:
\begin{equation}
Det_{Acc} = \frac{TPR^{(k)} + TNR^{(u)}}{2}.
\end{equation}

\subsubsection{Competitor Selection}
In the experiments, we follow the following criteria to select the comparison methods: 1) All competitors are published or presented in the most recent years; 2) A wide range of techniques are covered including both traditional methods and NN-based methods; 3) They have stated or demonstrated the state-of-the-art in malware recognition domain; and 4) All competitors are evaluated fairly under the same criteria with their best (reported) results. 

\begin{figure*}[t]
  \centering
  \subfigure[Classification]{
    \label{grid_acc}
    \label{fig:subfig:a} 
    \includegraphics[width=2.2in]{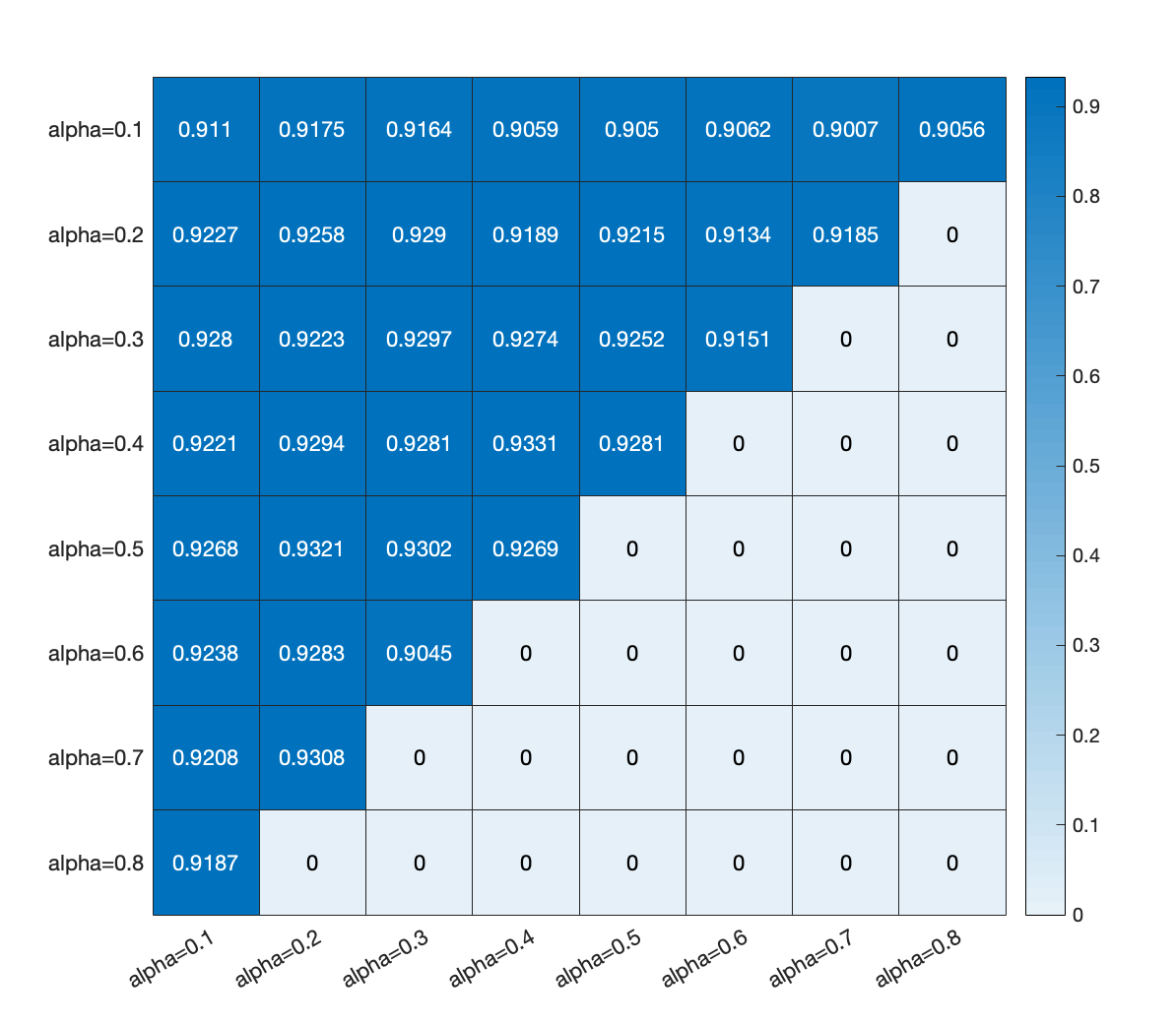}}
  \subfigure[Detection]{
    \label{grid_det}
    \label{fig:subfig:b} 
    \includegraphics[width=2.2in]{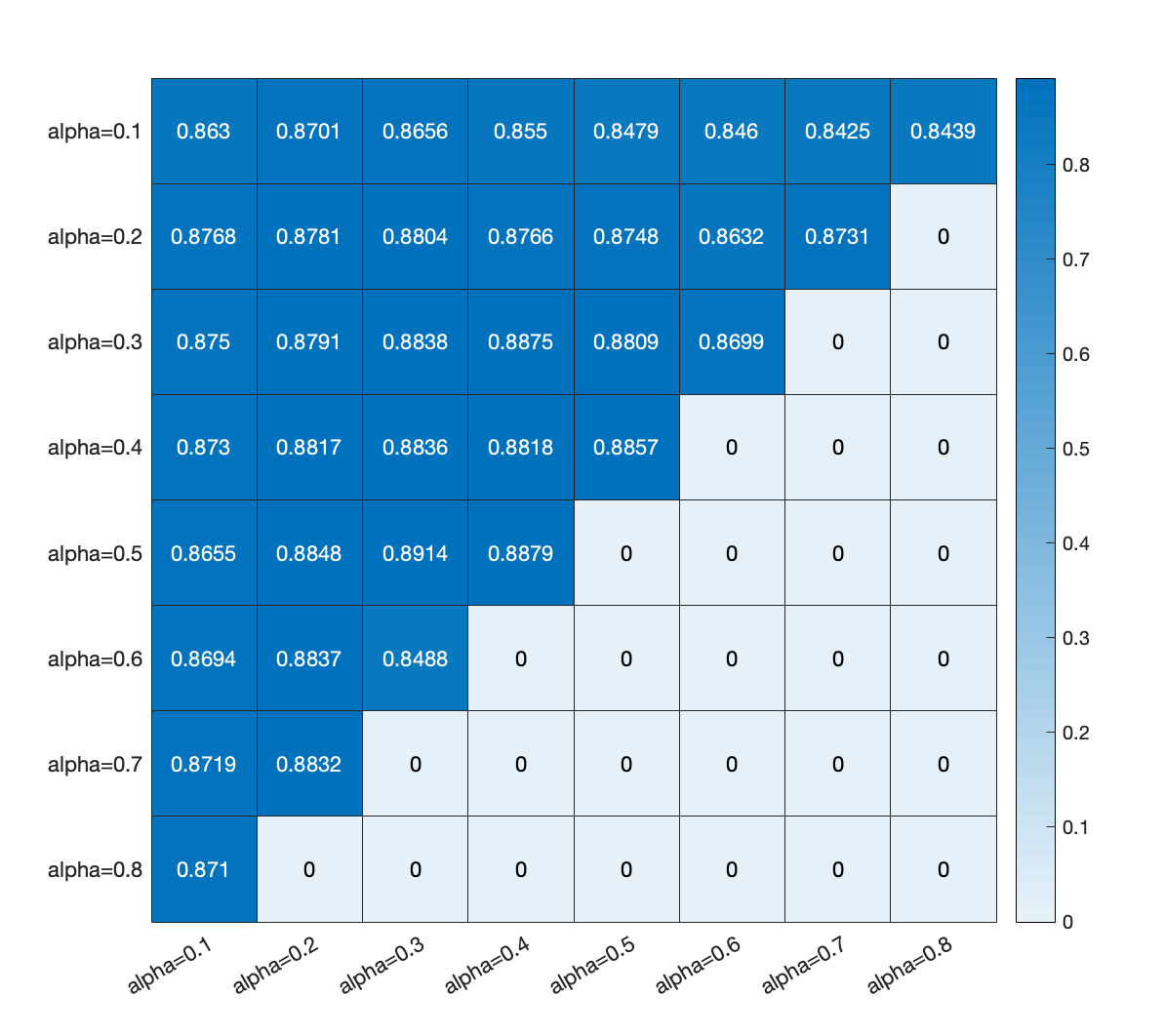}}
  \subfigure[Arithmetical Mean]{
    \label{grid_avg}
    \label{fig:subfig:b} 
    \includegraphics[width=2.2in]{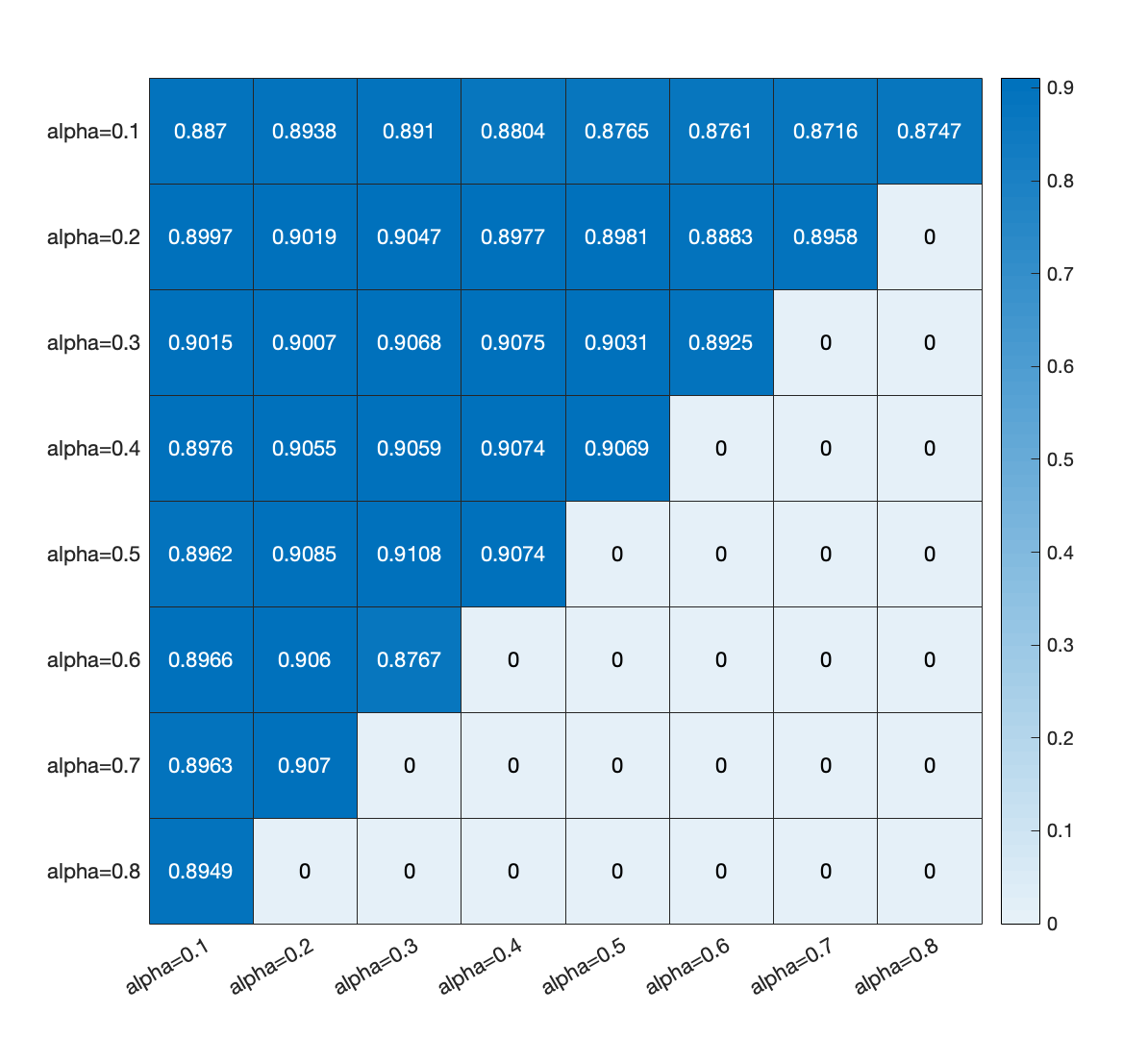}}
\caption{The sensibilities of hyper-parameters $\alpha$ and $\beta$ on the performance of (a) classification, (b) detection, and (c) arithmetical mean of both metrics. In each grid, the row denotes the setting of $\alpha$ and the column denotes the setting of $\beta$ (better viewed in color).}
  \label{grid}
\end{figure*}

\subsection{Implementation}
\subsubsection{Network and Training Details}
Our method is implemented by Pytorch and is trained on one NVIDIA RTX 3090 GPU. As the most essential building block in our \textit{MDENet}, the multi-modal encoders parallelly integrate a numeric encoder and textual encoder. 
Specifically, the numeric encoder is built upon a multi-scale pipeline, of which the global receptive module is achieved by two 1$\times$1 convolutions with 128 and 64 channels, respectively, wherein, a Gaussian function as stated in Eq.~(4) is used as the similarity metric. A shared convolutional neural network immediately follows afterwards for deeper processing. The local receptive module simply goes through one 1$\times$1 convolution with 64 channels and the above shared convolutional neural network to coordinate the training. The shared convolutional neural network consists of four convolutional layers with the pooling operations removed along with a larger kernel size and stride to fit our special design for malware. These two parts are concatenated and processed with a linear transformation. We briefly list the architecture in TABLE~\ref{network}. 
The textual encoder is implemented with pre-trained Transformer Encoder, i.e., borrows from \textit{BERT}~\cite{devlin2018bert}, with multi-head self-attention. The output of textual encoder is followed by a linear transformation with its output set to 1024. Last, the classifier is simply implemented with just one fully-connected layer.

\begin{table}[h]
\caption{Architecture of Numeric Encoder }
\label{network}    
\centering
\begin{tabular}{|ccc|ccc|}
\hline
\multicolumn{3}{|c|}{Global Receptive}                           & \multicolumn{3}{c|}{Local Receptive}                            \\ \hline
\multicolumn{3}{|c|}{1$\times$1 Conv (Channel=128) }                                & \multicolumn{3}{c|}{\multirow{3}{*}{1$\times$1 Conv (Channel=64)}}               \\ \cline{1-3}
\multicolumn{3}{|c|}{Gaussian Similarity}                               & \multicolumn{3}{c|}{}                                 \\ \cline{1-3}
\multicolumn{3}{|c|}{1$\times$1 Conv (Channel=64)}                             & \multicolumn{3}{c|}{}                                 \\ \hline
\multicolumn{6}{|c|}{Shared Convolutional Neural Network}                                                                                                            \\ \hline
\multicolumn{1}{|c|}{Channel}   & \multicolumn{1}{c|}{Kernel} & Stride & \multicolumn{1}{c|}{Padding} & \multicolumn{1}{c|}{ReLU} & BatchNorm \\ \hline
\multicolumn{1}{|c|}{64}  & \multicolumn{1}{c|}{5} & 1 & \multicolumn{1}{c|}{2} & \multicolumn{1}{c|}{$\checkmark$}  & $\checkmark$  \\ \hline
\multicolumn{1}{|c|}{128} & \multicolumn{1}{c|}{5} & 2 & \multicolumn{1}{c|}{2} & \multicolumn{1}{c|}{$\checkmark$}  & $\checkmark$  \\ \hline
\multicolumn{1}{|c|}{256} & \multicolumn{1}{c|}{5} & 2 & \multicolumn{1}{c|}{2} & \multicolumn{1}{c|}{$\checkmark$}  & $\checkmark$  \\ \hline
\multicolumn{1}{|c|}{512} & \multicolumn{1}{c|}{3} & 2 & \multicolumn{1}{c|}{1} & \multicolumn{1}{c|}{$\checkmark$}  & $\checkmark$  \\ \hline
\multicolumn{3}{|c|}{Linear Transformation $\to$ 512}                               & \multicolumn{3}{c|}{Linear Transformation $\to$ 512}                               \\ \hline
\multicolumn{6}{|c|}{Concatenation $\to$ 1024}                                                                                                            \\ \hline
\end{tabular}
\end{table}

\subsubsection{Hyper-parameters}
In Eq.~(20), we use three positive hyper-parameters $\alpha$, $\beta$, and $(1-\alpha-\beta)$ to balance $\operatorname{Loss}_{Cls}$, $\operatorname{Loss}_{Disc}$, and $\operatorname{Loss}_{Excl}$ in the overall objective. To determine the relative importance of each sub-loss function, we conduct a simple grid search to evaluate their  sensibilities. Specifically, we constrain $\alpha$, $\beta$, and $(1-\alpha-\beta)$ are no less than 0.1, and set an interval of 0.1 for them. Since their summation is also limited to 1 as a constrain for Eq.~(20), we thus can obtain a partially filled 8$\times$8 grid with 36 $(\alpha / \beta)$-pairs. 
We run our model with this grid on the proposed \textit{MAL-100$^{+}$} dataset with 500 epochs, and record the testing accuracy, i.e., $Cls_{Acc}$ and $Det_{Acc}$, for each $(\alpha / \beta)$-pair. The result of the grid search is demonstrated in Fig.~\ref{grid}. 
It can be observed that, the top-3 $(\alpha / \beta)$-pairs for classification are 0.4/0.4, 0.2/0.5, and 0.2/0.7. For detection, the top-3 $(\alpha / \beta)$-pairs are 0.3/0.5, 0.4/0.5, and 0.4/0.3. For arithmetical mean of both metrics, the top-3 $(\alpha / \beta)$-pairs are 0.3/0.5, 0.2/0.5, and 0.4/0.3. To balance the performance, we choose 0.3/0.5 as the $(\alpha / \beta)$-pair in our method.
During training, we use Adam~\cite{kingma2014adam} to optimize our method with a learning rate set as 0.0001 and a batch size set as 32. 
In Eq.~(19), since the positive hyper-parameter $\lambda$ is not a critical factor in this method, we empirically set it as 10. 
Last, hyper-parameter $\delta$ in Eq.~(22) is a calculable parameter relies on the data, which can be obtained by Eq.~(23). 


\subsection{Evaluation on \textit{Mailing}}

\subsubsection{Competitors}
We compare our proposed method, i.e., \textit{MDENet}, with 17 competitors evaluated on \textit{Mailing}, including 7 traditional methods as 
Nataraj~\textit{et al.}~\cite{nataraj2011malware}, 
Kalash~\textit{et al.}~\cite{kalash2018malware}, 
Yajamanam~\textit{et al.}~\cite{yajamanam2018deep}, 
Burnaev~\textit{et al.}~\cite{burnaev2016one}, 
\textit{GIST+SVM}~\cite{cui2018detection}, 
\textit{GLCM+SVM}~\cite{cui2018detection}, 
and \textit{EVM}~\cite{rudd2018extreme}; 
and 11 DNNs-based methods as
SoftMax-DNNs,  
\textit{VGG-VeryDeep-19}~\cite{yue2017imbalanced}, 
Cui~\textit{et al.}~\cite{cui2018detection}, 
\textit{IDA+DRBA}~\cite{cui2018detection}, 
\textit{NSGA-II}~\cite{cui2019malicious}, 
\textit{MCNN}~\cite{kalash2018malware}, 
\textit{IMCFN}~\cite{vasan2020imcfn}, 
\textit{OpenMax}~\cite{bendale2016towards}, 
\textit{tGAN}~\cite{kim2017malware}, 
\textit{tDCGAN}~\cite{kim2018zero}, 
and Guo~\textit{et al.}~\cite{guo2021conservative}. 

\subsubsection{Results and Analysis}
The evaluation results on \textit{Mailing} dataset are demonstrated in TABLE~\ref{table_mailing}. We can observe that our method outperforms all competitors with a classification accuracy as 99.32\%, and a detection accuracy as 90.77\%. More specifically, although \textit{Mailing} is a simple dataset for malware recognition task and all competitors have achieved fairly good performance, we can still find that our method further widens this gap especially for detection, where the margin is from 1.87\% to 38.05\%. Moreover, we can also observe that the DNNs-based methods usually perform much better than traditional methods due to the more complex modeling ability of deep models. 
It it noticed that despite our splitting uses fewer malware samples for the train-set, i.e., 6,400 out of total 9,339 samples, our method can still achieve the best results which fully verifies the effectiveness.

\begin{table*}[t]
\centering
\setlength{\tabcolsep}{6.0mm}{
\begin{threeparttable}  
\caption{Results on Mailing Dataset}  
\label{table_mailing}    
\begin{tabular}{lcccccccc}  
\toprule  
\multirow{2}{*}{Method}
&\multirow{2}{*}{Category}
&\multicolumn{2}{c}{Known}
&\multicolumn{1}{c}{Unknown}
&\multicolumn{1}{c}{Classification}
&\multicolumn{1}{c}{Detection}\cr  

\cmidrule(lr){3-4} 
&    &Train \#  &Test \#             &Test \#            &$Cls_{Acc}$ (\%)      &$Det_{Acc}$ (\%)  \cr  
\midrule  
Nataraj \textit{et al.} \cite{nataraj2011malware}     &TM         &8,394       &945     & -         &97.18      & -  \cr
Burnaev \textit{et al.} \cite{burnaev2016one}     &TM         &6,400       &1,617     &1,322         &94.19      &52.72      \cr
Kalash \textit{et al.} \cite{kalash2018malware}     &TM         &8,394       &945     & -         &93.23      & -      \cr
Yajamanam \textit{et al.} \cite{yajamanam2018deep}    &TM      &8394       &945     & -         &97.00      & -    \cr
\textit{GIST+SVM} \cite{cui2018detection}    &TM         &8,394       &945     & -         &92.20      & -      \cr
\textit{GLCM+SVM} \cite{cui2018detection}    &TM         &8,394       &945     & -         &93.20      & -      \cr
\textit{EVM} \cite{rudd2018extreme}    &TM         &6,400       &1,617     &1,322         &98.40      &81.70      \cr
\textit{SoftMax-DNNs}     &DNNs         &6,400       &1,617     &1,322         &98.08      &73.45   \cr
\textit{tGAN} \cite{kim2017malware}    &DNNs         &8,394       &945     & -        &96.82      & -      \cr
\textit{VGG-VeryDeep-19} \cite{yue2017imbalanced}    &DNNs         &5,603       &1,868     & -         &97.32      & -      \cr
Cui \textit{et al} \cite{cui2018detection}    &DNNs         &8,394       &945     & -         &97.60      & -      \cr
\textit{OpenMax} \cite{bendale2016towards}    &DNNs         &6,400       &1,617     &1,322         &98.70      &84.34      \cr
\textit{IDA+DRBA} \cite{cui2018detection}    &DNNs         &8,394       &945     & -         &94.50      & -      \cr
\textit{NSGA-II} \cite{cui2019malicious}    &DNNs         &8,394       &945     & -         &97.60      & -      \cr
\textit{MCNN} \cite{kalash2018malware}     &DNNs         &8,394       &945     & -         &98.52      & -      \cr
\textit{tDCGAN} \cite{kim2018zero}    &DNNs         &8,394       &945     & -        &97.66      & -      \cr
\textit{IMCFN} \cite{vasan2020imcfn}    &DNNs         &6,537       &2,802     & -         &98.82      & -      \cr
Guo~\textit{et al.}~\cite{guo2021conservative}    &DNNs       &6,400       &1,617     &1,322         &99.01      &88.90      \cr

\textbf{\textit{MDENet (ours)}}    &DNNs       &6,400       &1,617     &1,322         &\textbf{99.32}      &\textbf{90.77}      \cr
\bottomrule  
\end{tabular} 
\footnotesize{In `Category' column, `TM' denotes the traditional method and `DNNs' denotes the deep neural networks-based method. \# is the number of samples, and '-' indicates that there is no reported result associated with this method.}
\end{threeparttable}
}
\end{table*}

\subsection{Evaluation on \textit{MAL-100$^{+}$}}

\subsubsection{Competitors}
For the proposed malware dataset \textit{MAL-100$^{+}$}, we compare our method with 7 competitors including 2 traditional method as Burnaev~\textit{et al.}~\cite{burnaev2016one} and \textit{EVM}~\cite{rudd2018extreme}; 
and 5 DNNs-based methods as 
\textit{OpenMax}~\cite{bendale2016towards}, 
\textit{SoftMax-DNNs}, 
\textit{tGAN}~\cite{kim2017malware}, 
\textit{tDCGAN}~\cite{kim2018zero}, 
and Guo~\textit{et al.}~\cite{guo2021conservative}. 

\subsubsection{Results and Analysis}
The evaluation results on \textit{MAL-100$^{+}$} dataset are demonstrated in TABLE~\ref{table_mal100}. It can be observed that our method significantly expands its leading role against all competitors in both classification and detection. Specifically, the classification and detection accuracy achieves 94.30\% and 90.40\%. The improved margin is around $[3.13\%, 61.74\%]$ and $[4.17\%, 37.82\%]$, respectively. Among them, the traditional method, e.g., Burnaev~\textit{et al.}~\cite{burnaev2016one}, obtains a relatively acceptable detection accuracy as 52.58\%, which contrasts to its poor performance in classification. Again, the DNNs-based methods, i.e., 
\textit{OpenMax}~\cite{bendale2016towards}, 
\textit{SoftMax-DNNs}, 
\textit{tGAN}~\cite{kim2017malware}, 
\textit{tDCGAN}~\cite{kim2018zero}, Guo~\textit{et al.}~\cite{guo2021conservative}, 
and our \textit{MDENet}, perform much better than traditional methods. 
Notably, since the utilization of multi-modalities and dual-embedding space learning in our method, we significantly improve the recognition performance against the strongest competitor, i.e., Guo~\textit{et al.}~\cite{guo2021conservative}, and result in larger margins in \textit{MAL-100$^{+}$}. 

\begin{table*}[t]
\centering
\setlength{\tabcolsep}{6.0mm}{
\begin{threeparttable}  
\caption{Results on MAL-100$^{+}$}  
\label{table_mal100}    
\begin{tabular}{lcccccccc}  
\toprule  
\multirow{2}{*}{Method}
&\multirow{2}{*}{Category}
&\multicolumn{2}{c}{Known}
&\multicolumn{1}{c}{Unknown}
&\multicolumn{1}{c}{Classification}
&\multicolumn{1}{c}{Detection}\cr  

\cmidrule(lr){3-4} 
&    &Train \#  &Test \#             &Test \#            &$Cls_{Acc}$ (\%)      &$Det_{Acc}$ (\%)  \cr  
\midrule
Burnaev \textit{et al.} \cite{burnaev2016one}     &TM         &31,523       &7,823     &17,135         &32.56      &52.58      \cr
\textit{EVM} \cite{rudd2018extreme}    &TM         &31,523       &7,823     &17,135         &84.10      &68.69      \cr
\textit{SoftMax-DNNs}     &DNNs         &31,523       &7,823     &17,135         &76.45      &63.02   \cr  
\textit{OpenMax} \cite{bendale2016towards}    &DNNs         &31,523       &7,823     &17,135         &85.38      &70.57      \cr
\textit{tGAN} \cite{kim2017malware}    &DNNs         &31,523       &7,823     &17,135         &77.23      &65.25      \cr
\textit{tDCGAN} \cite{kim2018zero}    &DNNs         &31,523       &7,823     &17,135         &81.45      &67.04      \cr
Guo~\textit{et al.}~\cite{guo2021conservative}    &DNNs        &31,523       &7,823     &17,135         &91.17      &86.23      \cr

\textbf{\textit{MDENet (ours)}}    &DNNs       &31,523       &7,823     &17,135         &\textbf{94.30}      &\textbf{90.40}      \cr
\bottomrule  
\end{tabular} 
\footnotesize{In `Category' column, `TM' denotes the traditional method and `DNNs' denotes the deep neural networks-based method. \# is the number of samples, and '-' indicates that there is no reported result associated with this method.}
\end{threeparttable}
}
\end{table*}

\subsection{Ablation Analysis}
We conduct an ablation analysis (TABLE \ref{ablation_study}) on our method to further investigate the influence of multi-modal information, namely, the numeric information and the textual information, of the extracted malware features in \textit{MAL-100$^{+}$}. 
The performance is compared in three scenarios: 1) Only the malware images, i.e., numeric information, are used; 2) Only the malware sentences, i.e., textual information, are used; and 3) Both of them are used as a multi-modal manner. 
%
%

First, in scenarios 1), our method can outperform all other representative methods only except for Guo~\textit{et al.}~\cite{guo2021conservative} (TABLE~\ref{table_mal100}) by using only the malware images, which indicates that the dual-embedding space learning can result in more diversities in the obtained representation vectors and thus improve the recognition performance. Notably, Guo~\textit{et al.}~\cite{guo2021conservative} is only slightly better in the classification task (0.07\%) than scenario 1) of our method, while its computational overhead is significantly larger due to the complex training of GANs~\cite{goodfellow2014generative}. Moreover, our detection performance is constantly better than all competitors including Guo~\textit{et al.}~\cite{guo2021conservative}, which suggests our distance-based detection mechanism (sect.~\ref{Distance-based Detection for Unknown Families}) can well mitigate the detection degradation. 
Second, in scenarios 2), to use only the malware sentences can improve neither classification nor detection accuracy since the $Cls_{Acc}$ and $Det_{Acc}$ are lower than most competitors. 
Despite all that, it should be also noted that the performance is still better than some traditional methods, e.g, Burnaev~\textit{et al.}~\cite{burnaev2016one}, which indicates that the textual features also contain useful information.  
Last, in scenarios 3), if both the malware images and sentences are used, our method can achieve the best performance in both classification and detection tasks. The ablation analysis fully demonstrates the utility and effectiveness of our method.

\begin{table}[t]
\centering
\setlength{\tabcolsep}{4.7mm}{
\begin{threeparttable}  
\caption{Ablation Study of the Proposed Method}
\label{ablation_study}      
\begin{tabular}{cccccc}  
\toprule  
\multicolumn{2}{c}{Modality}
&\multicolumn{1}{c}{Classification}
&\multicolumn{1}{c}{Detection}\cr  

\cmidrule(lr){1-2} 
Image    &Sentence &$Cls_{Acc}$ (\%)  &$Det_{Acc}$ (\%)      \cr  
\midrule  
$\checkmark$   &    &91.10  &87.40       \cr
   &$\checkmark$    &73.07  &56.91       \cr
$\checkmark$   &$\checkmark$  &94.30  &90.40       \cr
\bottomrule  
\end{tabular}
\footnotesize{The Malware image and sentence are two modalities of the numeric and textual information, respectively.} 
\end{threeparttable}
}
\end{table} 

\begin{figure*}[t]
  \centering
  \subfigure[Classification confusion matrix (ours)]{
    \label{cm:a} 
    \includegraphics[width=2.3in]{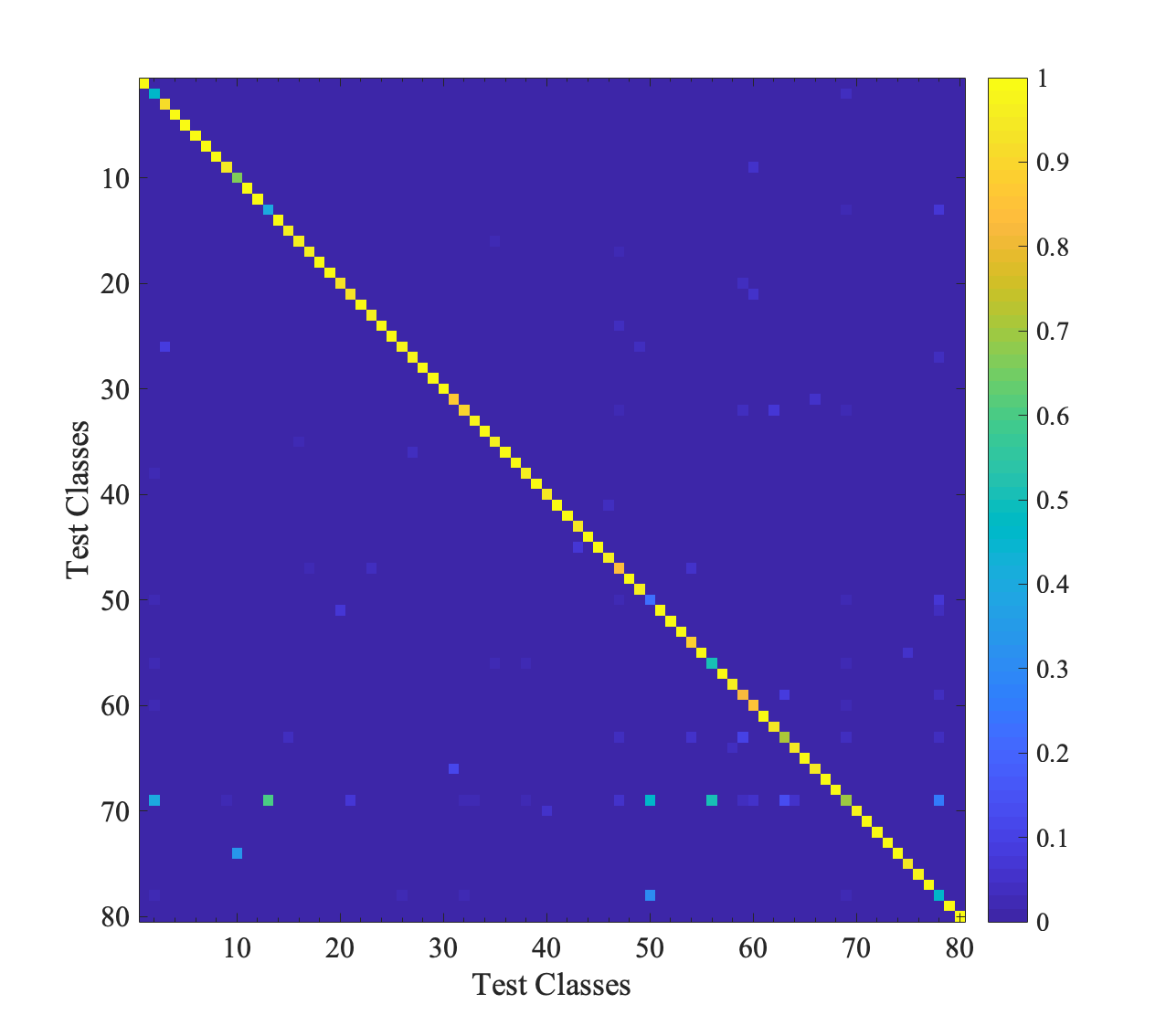}}
  \subfigure[Landscape of the confusion matrix (ours)]{
    \label{cm:b} 
    \includegraphics[width=2.3in]{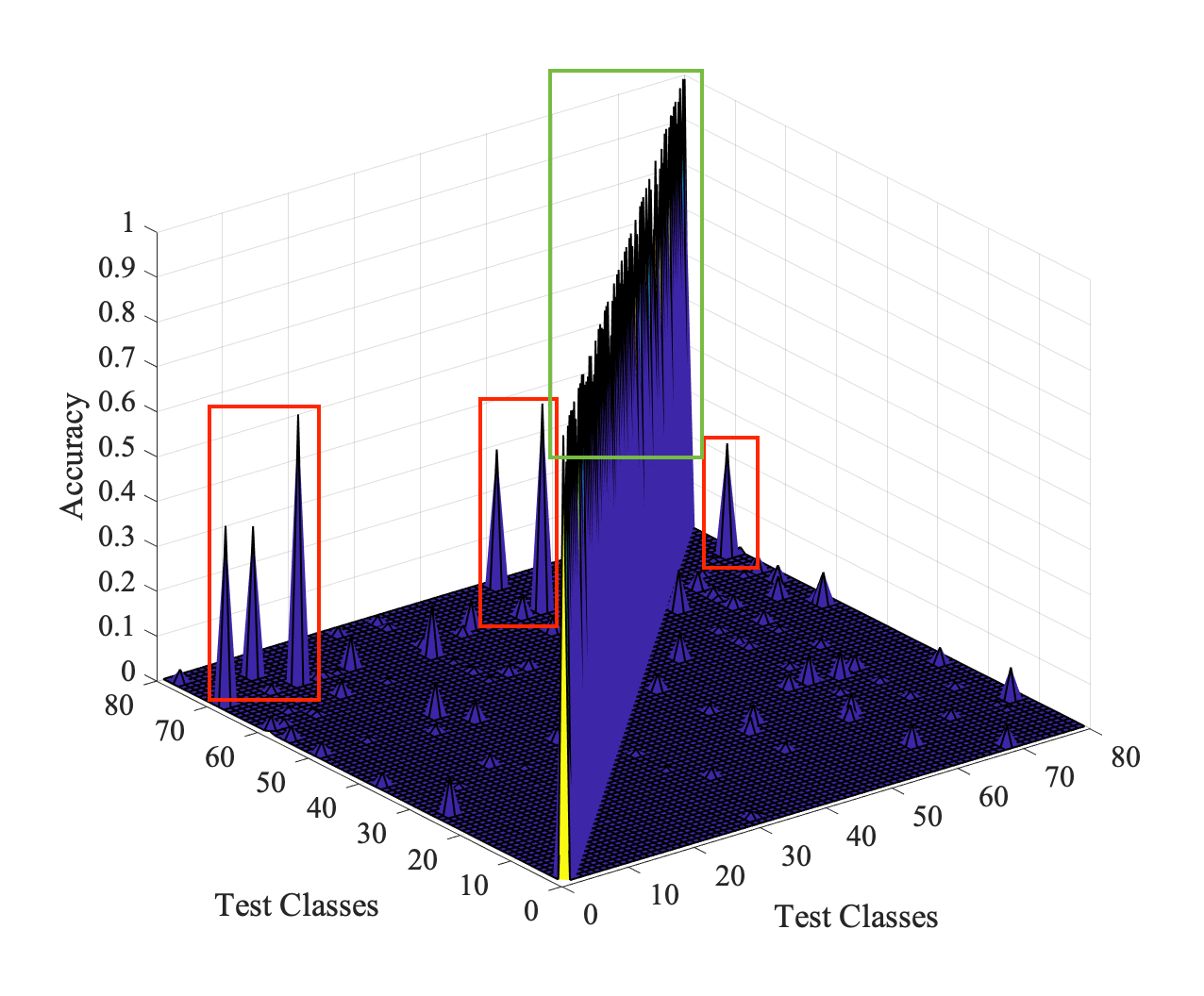}}
  \subfigure[Landscape of the confusion matrix (Guo~\textit{et al.}~\cite{guo2021conservative})]{
    \label{cm:c} 
    \includegraphics[width=2.3in]{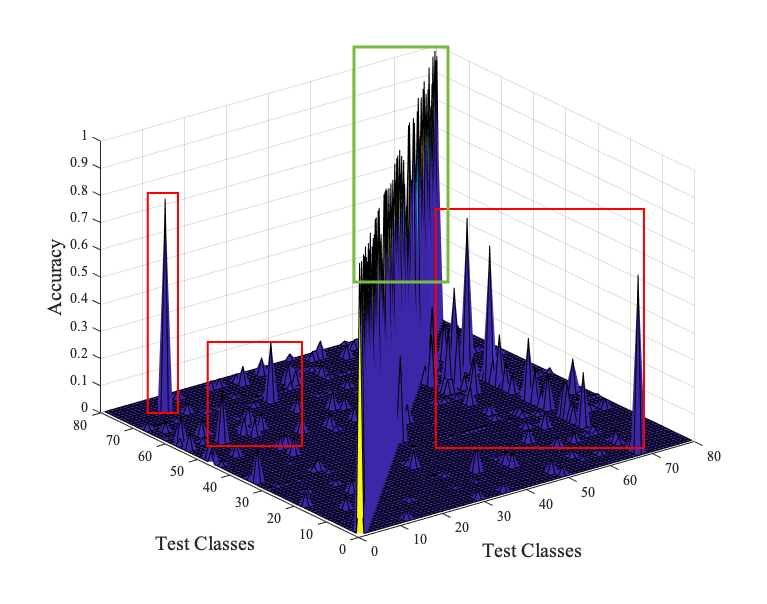}}
\caption{Per-family classification performance of the proposed \textit{MDENet} on \textit{MAL-100$^{+}$}: (a) the classification confusion matrix and (b) the landscape of the the confusion matrix. Moreover, we also demonstrate the landscape of the current state-of-the-art competitor, i.e., Guo~\textit{et al.}~\cite{guo2021conservative}, in (c)  
(better viewed in color and zoom-in mode).}
  \label{CM}
\end{figure*}

\subsection{Fine-grained Accuracy}
We evaluate the fine-grained per-family classification performance to further demonstrate the predictive power of our method. Specifically, we record the prediction results for each testing malware families to construct a confusion matrix on \textit{MAL-100$^{+}$} with a full training process. 
As demonstrate in Fig.~\ref{cm:a} and Fig.~\ref{cm:b}, we show the confusion matrix and the corresponding landscape for each testing families of our method, wherein, the column position denotes the ground truth, and the row position denotes the predicted results. Thus, the diagonal position can calculate the per-family classification accuracy easily. Moreover, we also demonstrate the landscape of the current state-of-the-art competitor, i.e., Guo~\textit{et al.}~\cite{guo2021conservative}, in Fig.~\ref{cm:c}.
It can be observed that our method can constantly obtain good performance along with better fine-grained accuracy for each malware family, and thus contributes to the better overall accuracy. 
Notably, from Fig.~\ref{cm:b} and Fig.~\ref{cm:c}, it can also be observed that compared with the current state-of-the-art competitor, i.e., Guo~\textit{et al.}~\cite{guo2021conservative}, our method is capable of predicting much more balanced per-family accuracy (covered by \textit{green} rectangle) and fewer mis-classified malware families (covered by \textit{red} rectangles), which can demonstrate that the proposed \textit{MDENet} is more robust to real-world prediction and better handle some hard families/examples.

\section{Conclusion}
\label{sec:conclusion}
In this paper, we proposed a novel and robust malware open-set recognition framework, dubbed \textit{MDENet}, involving the multi-modal encoders and dual-embedding space learning approach. Our method can fully make use of multi-modal malware features and fuse them into an enhanced malware feature space that with more diversities. A distance-based detection mechanism is proposed to further mitigate the detection degradation. Moreover, we also contribute a new version of large-scale malware dataset \textit{MAL-100$^{+}$}. Experimental results verified the effectiveness of our method.

\bibliographystyle{IEEEtran}
\bibliography{malware_osr}
\end{document}